\documentclass{aa} % printer
\usepackage{graphicx}
\usepackage{txfonts}
\usepackage{natbib}
\usepackage{float}

\begin{document}

\title{Gas structure inside dust cavities of transition disks: \mbox{Oph IRS 48 observed by ALMA}}

\author{Simon Bruderer\inst{\ref{inst:mpe}}, Nienke van der Marel\inst{\ref{inst:leiden}}, Ewine F. van Dishoeck\inst{\ref{inst:leiden},\ref{inst:mpe}}, Tim A. van Kempen\inst{\ref{inst:leiden}}}

\institute{
Max-Planck-Institut f\"{u}r Extraterrestrische Physik, Giessenbachstrasse 1, 85748 Garching, Germany\label{inst:mpe}
\and
Leiden Observatory, Leiden University, P.O. Box 9513, 2300 RA Leiden, The Netherlands\label{inst:leiden}
}

\date{Accepted by A\&A, December 10th 2013}

\titlerunning{The gas structure inside dust cavities of the transition disks}
\authorrunning{S. Bruderer et al.}

\offprints{Simon Bruderer,\\ \email{simonbruderer@gmail.com}}

\abstract
{Transition disks are recognized by the absence of emission of small dust grains inside a radius of up to several 10s of AUs. Due to the lack of angular resolution and sensitivity, the gas content of such dust holes has not yet been determined, but is of importance to constrain the mechanism leading to the dust holes. Transition disks are thought to currently undergo the process of dispersal, setting an end to the giant planet formation process.} % Context
{We present new high-resolution observations with the Atacama Large Millimeter/submillimeter Array (ALMA) of gas lines towards the transition disk Oph IRS 48 previously shown to host a large dust trap. ALMA has detected the $J=6-5$ line of $^{12}$CO and C$^{17}$O around 690 GHz (434 $\mu$m) at a resolution of $\sim$0.25$''$ corresponding to $\sim$30 AU (FWHM). The observed gas lines are used to set constraints on the gas surface density profile.} % Aims
{New models of the physical-chemical structure of gas and dust in Oph IRS 48 are developed to reproduce the CO line emission together with the spectral energy distribution (SED) and the VLT-VISIR 18.7 $\mu$m dust continuum images. Integrated intensity cuts and the total spectrum from models having different trial gas surface density profiles are compared to observations. The main parameters varied are the drop of gas surface density inside the dust free cavity with a radius of $60$ AU and inside the gas depleted innermost 20 AU. Using the derived surface density profiles, predictions for other CO isotopologues are made, which can be tested by future ALMA observations of the object.} % Method
{From the ALMA data we find a total gas mass of the disk of $1.4 \times 10^{-4}$ M$_\odot$. This gas mass yields a gas-to-dust ratio of $\sim$10, but with considerable uncertainty. Inside 60 AU, the gas surface density drops by a factor of $\sim$12 for an assumed surface density slope of $\gamma=1$ ($\Sigma \propto r^{-\gamma}$). Inside 20 AU, the gas surface density drops by a factor of at least 110. The drops are measured relative to the extrapolation to small radii of the surface density law at radii $>$60 AU. The inner radius of the gas disk at 20 AU can be constrained to better than $\pm$5 AU.} % Results
{The derived gas surface density profile points to the clearing of the cavity by one or more massive planet/companion rather than just photoevaporation or grain-growth.} % Conclusions

\maketitle

%%%%%%%%%%%%%%%%%%%%%%%%%%%%%%%%%%%%%%%%%%%%%%%%%%%%%%%%%%%%%%%%%%%%%%%%%%%%%%%%%%%%%%%%%%%%%%%%%
%
% Introduction
%
%%%%%%%%%%%%%%%%%%%%%%%%%%%%%%%%%%%%%%%%%%%%%%%%%%%%%%%%%%%%%%%%%%%%%%%%%%%%%%%%%%%%%%%%%%%%%%%%%
%%%%%%%%%%%%%%%%%%%%%%%%%%%%%%%%%%%%%%%%%%%%%%%%%%%%%%%%%%%%%%%%%%%%%%%%%%%%%%%%%%%%%%%%%%%%%%%%%

\section{Introduction} \label{sec:intro}

Protoplanetary disks are the cradle for young planets. As the disk evolves from a gas-rich T Tauri disk to a gas-poor debris disk, the disk mass steadily decreases (\citealt{Armitage11} for a review). The dispersal of the disk sets an end to giant planet formation and thus determines the time-scale of the planet formation process. Detailed understanding of the disk evolution and in particular of the disk dispersal is thus crucial for our understanding how, where, and when planets form. A particular class of protoplanetary disks, called transition disks, is thought to be currently in the stage of dispersal. The dust distribution in transition disks has been studied thoroughly through the spectral energy distribution (SED) from ultraviolet (UV) to millimeter wavelength and in continuum imaging, but still very little is known about the presence and characteristics of the gas in these disks. Here, we use spatially resolved high signal-to-noise gas line observations towards a transition disk to derive the gas structure, and, implicitly determine the origin of the dust holes in this disk.

Transition disks are commonly identified through their SED showing a strong excess over the stellar photosphere at wavelengths $\gtrsim$20$\mu$m, but little excess at shorter wavelengths. The deficit of near infrared excess arises from the absence of hot small dust close to the star, suggesting the presence of an inner dust cavity (\citealt{Strom89,Calvet02,Brown07}). Several of these dust holes have been imaged directly in (sub)millimeter interferometric imaging (\citealt{Pietu06,Brown09,Andrews11,Isella12,Isella13}). Recently, candidates for young planets have been found in cavities of the transition disks \object{T Cha} by \cite{Huelamo11}, \object{LkCa 15} by \cite{Kraus12}, and \object{HD 100546} by \cite{Quanz13}.

The gas component studied in this work is key to distinguish the different mechanisms proposed for the formation of the dust cavity: grain growth, photo-evaporation and clearing by a planet or substellar companion. Grain growth does not affect the gas density, while photo-evaporation removes gas and dust simultaneously. Clearing by a planet or substellar companion reduces the amount of gas in the cavity depending on the mass of the companion and other parameters (e.g. viscosity, e.g. \citealt{Zhu11,DodsonRobinson11,Pinilla12b,Mulders13,Fung13}). The planet/companion-disk interaction can also induce perturbations in the gas structure leading to local gas pressure maxima. These pressure maxima can prevent large ($\gtrsim$ mm-sized) dust grains from quickly drifting towards the star before planet formation through dust coagulation and core accretion can take place (e.g. \citealt{Whipple72,Rice06,Alexander07,Garaud07,Kretke07,Dzyurkevich10,Pinilla12b,Birnstiel13,Lyra13}). Thus, the gas structure in transition disks is of direct importance for planet formation. Moreover, the magnitude of any drop in surface density profile is directly related to the mass of the companion.

Gas line observations towards transition disks have been carried out at near infrared and submillimeter wavelengths. The CO rovibrational emission line at 4.7 $\mu$m tracing the several 100 K warm gas from the inner regions of the disk has been detected for several transition disks  (\citealt{Goto06,Pontoppidan08,VanderPlas09,Salyk09,Brown12,Brown13}). In contrast, submillimeter observations of the rotational gas lines trace the bulk of the colder gas and are more suitable for constraining the gas mass and distribution inside the cavity (\citealt{Bruderer13}). So far, submillimeter observations have suffered from the low angular resolution and sensitivity, barely detected the gas in the outer disk, and did not allow the study of properties of the gas inside the cavity (\citealt{Dutrey08,Lyo11}). With the Atacama Large Millimeter/submillimeter Array (ALMA\footnote{http://www.almaobservatory.org}) the cold gas in the inner regions of transition disks can be imaged for the first time. In this paper, we present a detailed analysis of the gas distribution in the transition disk around the young Herbig star \object{Oph IRS 48}, using spatially resolved submillimeter ALMA Band 9 (690 GHz) observations.

Oph IRS 48 ($\alpha_{\rm 2000}=$16$^{\rm h}$27$^{\rm m}$37\fs18, $\delta_{\rm 2000}=$-24\degr 30\arcmin 35.3\arcsec) is an A0 star located in the $\rho$ Ophiuchi star formation region at a distance of 120 parsec (\citealt{Loinard08,Brown12}). The star shows weak accretion signatures (\citealt{Salyk13}). A ring-like structure peaking at $\sim$60 AU radius was found by spatially resolved 18.7 $\mu$m imaging of the dust continuum (\citealt{Geers07}). IRS 48 shows very bright polycyclic aromatic hydrocarbon (PAH) emission centered on the star within this hole. VLT-CRIRES spectra of the 4.7 $\mu$m CO line reveal a gas ring with a radius of $\sim$25-35 AU \citep{Brown12}. The submillimeter continuum (685 GHz or 0.43 mm) of our ALMA observations was presented in \citet{vdMarel13}. In contrast to the gas and small dust grains, the millimeter dust is concentrated on one side of the disk, with a high azimuthal contrast of $>$130 compared to the other side. This asymmetric dust distribution was interpreted as a dust trap, triggered by the presence of a substellar companion inside $\sim$20 AU. 

In this work we analyze the gas distribution of IRS 48 in detail, using the ALMA observations of $^{12}$CO $J=6-5$ and C$^{17}$O $J=6-5$ taken simultaneously with the continuum observations. Using the combined physical-chemical model by \citet{Bruderer12} and \citet{Bruderer13}, we derive constraints on the drop of the gas surface density profile at a radius of 60 AU, which is the outer radius of the dust cavity, and at $\sim$20 AU, where \citet{vdMarel13} have identified a gas hole possibly related to a companion situated inside. With the detailed modeling of IRS 48, we also introduce a framework to analyze upcoming ALMA observations of transition disks.

The paper is structured as follows: In Section \ref{sec:obs} we discuss the observations and data reduction. Section \ref{sec:results} presents spectra, integrated intensity maps and channel maps. In Section \ref{sec:analysis}, we present detailed models of IRS 48. We first discuss the dust structure derived from the spectral energy distribution and VISIR images and then compare models with different gas surface density profiles to the ALMA observations. Implications of the derived gas mass and gas surface density structure are given in Section \ref{sec:discussion}. The paper ends with a conclusions section. 

%%%%%%%%%%%%%%%%%%%%%%%%%%%%%%%%%%%%%%%%%%%%%%%%%%%%%%%%%%%%%%%%%%%%%%%%%%%%%%%%%%%%%%%%%%%%%%%%%
%
% Observations and data reduction
%
%%%%%%%%%%%%%%%%%%%%%%%%%%%%%%%%%%%%%%%%%%%%%%%%%%%%%%%%%%%%%%%%%%%%%%%%%%%%%%%%%%%%%%%%%%%%%%%%%
%%%%%%%%%%%%%%%%%%%%%%%%%%%%%%%%%%%%%%%%%%%%%%%%%%%%%%%%%%%%%%%%%%%%%%%%%%%%%%%%%%%%%%%%%%%%%%%%%

\section{Observations and data reduction} \label{sec:obs}

Band 9 ALMA Cycle 0 observations of IRS 48 were carried out in the extended configuration on June 6th and July 17th 2012 in three execution blocks of 1.7 hours each (one on June 6th and two on July 17th). In these blocks, 18 to 21 antennas were used with baselines up to 390 meter. The average precipitable water vapor (pwv) levels were 0.50, 0.34 and 0.17 mm, respectively. The spectral setup consisted of four spectral windows, centered at 674.00 , 678.84, 691.47 and 693.88 GHz, to target the C$^{17}$O $J=6-5$, CN $J=6_{11/2}-5_{11/2}$, $^{12}$CO $J=6-5$, and H$^{13}$CO$^+$ $J=8-7$ transitions. The CN and H$^{13}$CO$^+$ lines have however not been detected and are discussed in \citet{vdMarel13b} together with a detection of H$_2$CO $9_{18}-8_{17}$. The spectral windows consist of 3840 channels each with a channel separation of 488 kHz and thus a bandwidth of 1875 MHz. The final velocity resolution is 0.21 km s$^{-1}$ using a reference of 690 GHz and the rms noise level is 40 mJy beam$^{-1}$ channel$^{-1}$. The synthetic beam has a size of $0.32'' \times 0.21''$ ($38 \times 25$ AU) and a position angle of 96$^\circ$ (east-of-north).

Reduction and calibration of the data was performed using the Common Astronomy Software Application (CASA) version 3.4 and is further described in the Supplementary Online Material of \citet{vdMarel13}.

%%%%%%%%%%%%%%%%%%%%%%%%%%%%%%%%%%%%%%%%%%%%%%%%%%%%%%%%%%%%%%%%%%%%%%%%%%%%%%%%%%%%%%%%%%%%%%%%%
%
% Results
%
%%%%%%%%%%%%%%%%%%%%%%%%%%%%%%%%%%%%%%%%%%%%%%%%%%%%%%%%%%%%%%%%%%%%%%%%%%%%%%%%%%%%%%%%%%%%%%%%%
%%%%%%%%%%%%%%%%%%%%%%%%%%%%%%%%%%%%%%%%%%%%%%%%%%%%%%%%%%%%%%%%%%%%%%%%%%%%%%%%%%%%%%%%%%%%%%%%%

\section{Results} \label{sec:results}

\subsection{Integrated intensity maps} \label{sec:result_int}

\begin{figure*}[htb!]
\center
\includegraphics[width=1.0\hsize]{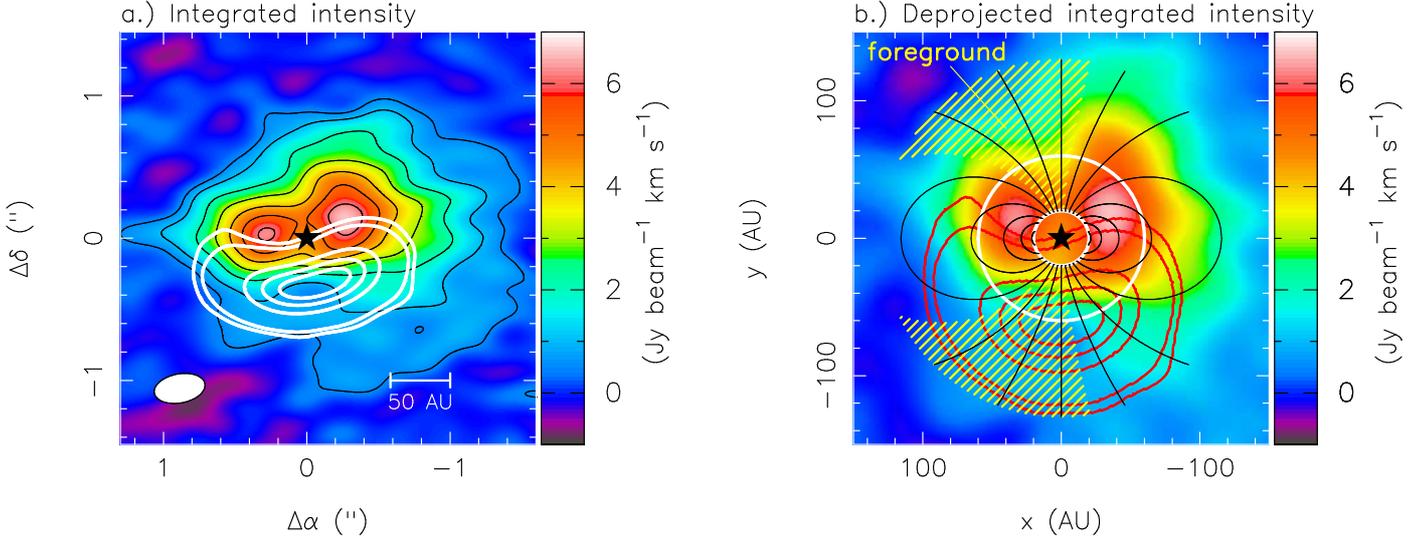}
\caption{\textit{a)} Continuum subtracted integrated intensity of $^{12}$CO $J=6-5$ in color. The black contour lines show $5\sigma, 10\sigma, 20\sigma, \ldots $ detections of the line ($1\sigma=73$ mJy beam$^{-1}$ km s$^{-1}$), white contour lines represent $5\sigma, 10\sigma, 100\sigma, 200\sigma, 300\sigma$ detections of the continuum ($1\sigma=0.82$ mJy beam$^{-1}$). The ALMA beam (FWHM) is indicated in the left lower corner. \textit{b)} Deprojected continuum subtracted integrated intensity maps of $^{12}$CO $J=6-5$ in color and the continuum with contours ($5\sigma, 10\sigma, 100\sigma, 200\sigma, 300\sigma$ detections). The ${\rm (x,y)}$-coordinates are along the major/minor axis of the observed disk. Thick white lines indicated radii of 20 and 60 AU and deprojected isovelocity contours of regions with velocity $\pm 1, \pm2, \ldots$ km s$^{-1}$ relative to ${\rm v}_{\rm lsr}$ towards us. The yellow shaded region indicates positions affected by foreground absorption.}
\label{fig:int_imgdeproj}
\end{figure*}

Figure \ref{fig:int_imgdeproj}a shows the continuum subtracted integrated intensity of the $^{12}$CO $J=6-5$ observation. The continuum emission at the same wavelength (430 $\mu$m) is indicated in white contour lines. The strongly lopsided shape of the continuum is interpreted by \citet{vdMarel13} as large $>$mm-size dust being trapped by a local pressure maximum due to a long-lived vortex in the gas, induced by a companion situated at a radius $<$20 AU. The integrated intensity of CO is elongated in the east-west direction and shows a drop at the stellar position. The stellar position (Table \ref{tab:basic_irs48}) is determined from the fastest velocity channels where CO emission has been detected. The elongation is due to the inclination of $i=50^\circ$. At the position of the continuum peak, CO emission is detected, but not particularly strong. This is however not inconsistent with a pressure maximum at this position, because $^{12}$CO is likely optically thick and a weak increase of density by less than a factor of 2 is sufficient to trigger dust trapping  (\citealt{Birnstiel13}; see further discussion in Section \ref{sec:disc_surfdusttrap}).

In Figure \ref{fig:int_imgdeproj}b, the continuum subtracted integrated intensity deprojected for inclination, position angle, and distance is shown. After deprojection, the disk is shown as seen face-on with coordinates ${\rm (x,y)}$ in AU along the major/minor axis of the observed disk. For the deprojection, the parameters given in Table \ref{tab:basic_irs48} are used. The inclination is determined by \cite{Geers07} from 18.7 $\mu$m dust continuum images to be $i=48\pm8^\circ$. This inclination is confirmed by \cite{Brown12} from CO ro-vibrational lines. We use $i=50^\circ$, which was found to agree best with the channel maps of $^{12}$CO (Figure \ref{fig:velo_chan}). \citet{Geers07} find a position angle of $98\pm3^\circ$ (east-of-north) from the 18.7 $\mu$m dust continuum images. This is consistent with ${\rm PA}=100.3^\circ$ determined from the position of the highest velocity channels where $^{12}$CO is detected (Figure \ref{fig:velo_chan}). We overlay deprojected isovelocity contour lines providing lines with the same velocity towards us. The contour lines are derived assuming a geometrically thin disk in Keplerian rotation around a 2 M$_{\odot}$ star (Table \ref{tab:basic_irs48}). The deprojected map shows weaker emission in the north-east compared to the north-west. The reason for this is absorption by a foreground cloud. The $\rho$ Ophiuchus region has several foreground layers and clouds with a high enough shielding to be abundant in CO (\citealt{Loren89,Boogert02,vanKempen09}). Such cold foreground CO can shield the disk emission. The foreground towards IRS 48 ($A_V$ = 11.5, \citealt{Brown12}) can provide enough line opacity ($\tau >$ few) in $^{12}$CO $J=6-5$ for a cloud with temperatures around 30 K and densities around $3 \times 10^4$ cm$^{-3}$. Foreground absorption usually affects only a narrow velocity range, since they are cold and have small intrinsic line widths. Towards IRS 48, velocities ${\rm v}_{\rm lsr}\sim$2.0 to $\sim$4.5 km s$^{-1}$ are affected indicated by the yellow shaded region (\citealt{vanKempen09}). 

Asymmetries in the east-west direction in the north of the disk ($y \gtrsim 50$ AU in the deprojected map) can thus be explained by foreground absorption. Between $y=-50$ AU and $y=50$ AU, the disk is less extended in the eastern direction, which cannot be explained by the foreground. An additional foreground layer at ${\rm v}_{\rm lsr}=1.5$ km s$^{-1}$ found towards Oph VLA 1623 by \citet{Murillo13b} would help to explain this asymmetry, but it is unclear if this previously unknown layer also extends to IRS 48. The projected distance of IRS 48 to VLA 1623 is $\sim$19$'$ or 0.7 pc. At $x \lesssim -20$ AU, an asymmetry in north-south direction is outside the region affected by foreground by more than the size of the beam of $20-30$ AU. For example at $x=-20$ AU, the emission at $y=-60$ AU is about a factor of 3 weaker compared to $y=60$ AU. At these positions the peak intensities are 1.7 Jy beam$^{-1}$ and 0.7 Jy beam$^{-1}$, which is stronger than the continuum peak of 0.3 Jy beam$^{-1}$. Thus, a pure line-to-continuum effect can also not explain this asymmetry. Since also the peak integrated intensity is slightly shifted to the north with respect to the major axis of the disk ($y=0$ AU), a north-south temperature gradient could be the reason for this asymmetry (see further discussion in Section \ref{sec:disc_surfdusttrap}). 

We conclude that some, but not all, of the asymmetries seen in the integrated intensity maps can be attributed to foreground absorption. Least affected by the foreground are regions close to the major axis of the disk.

\begin{table}[tbh]
\caption{Basic data of Oph IRS 48}
\label{tab:basic_irs48}
\centering
\begin{tabular}{lll}
\hline\hline
Parameter &  Value & Reference \\
\hline
Stellar position    & $\alpha_{\rm 2000}=$16$^{\rm h}$27$^{\rm m}$37\fs18  &  B13\\
                    &  $\delta_{\rm 2000}=$-24\degr 30\arcmin 35.3\arcsec  & \\
Distance            & 120 pc  & L08 \\
Inclination         & $i=50^\circ$ & G07, B13 \\
Systemic velocity   & v$_{\rm lsr}=4.55$ km s$^{-1}$ & vdM13 \\
Position angle      & ${\rm PA}=100.3^\circ$ (east-of-north) & G07, B13\\
Stellar type        & A0$^{+4}_{-1}$ & B12 \\
Stellar mass        & M$_* =$ 2 M$_\odot$ & B12 \\
Stellar luminosity  & 14.3 L$_\odot$ & B12 \\
Accretion rate      & $4 \times 10^{-9}$ M$_\odot$ yr$^{-1}$ & S13\\
\hline
\end{tabular}
\tablefoot{B13$=$this work, L08$=$\citet{Loinard08}, G07$=$\citet{Geers07}, B12$=$\citet{Brown12}, S13$=$\citet{Salyk13}, vdM13$=$\citet{vdMarel13}}
\end{table}

\subsection{Channel maps} \label{sec:result_chan}

\begin{figure*}[htb!]
\center
\includegraphics[width=1.0\hsize]{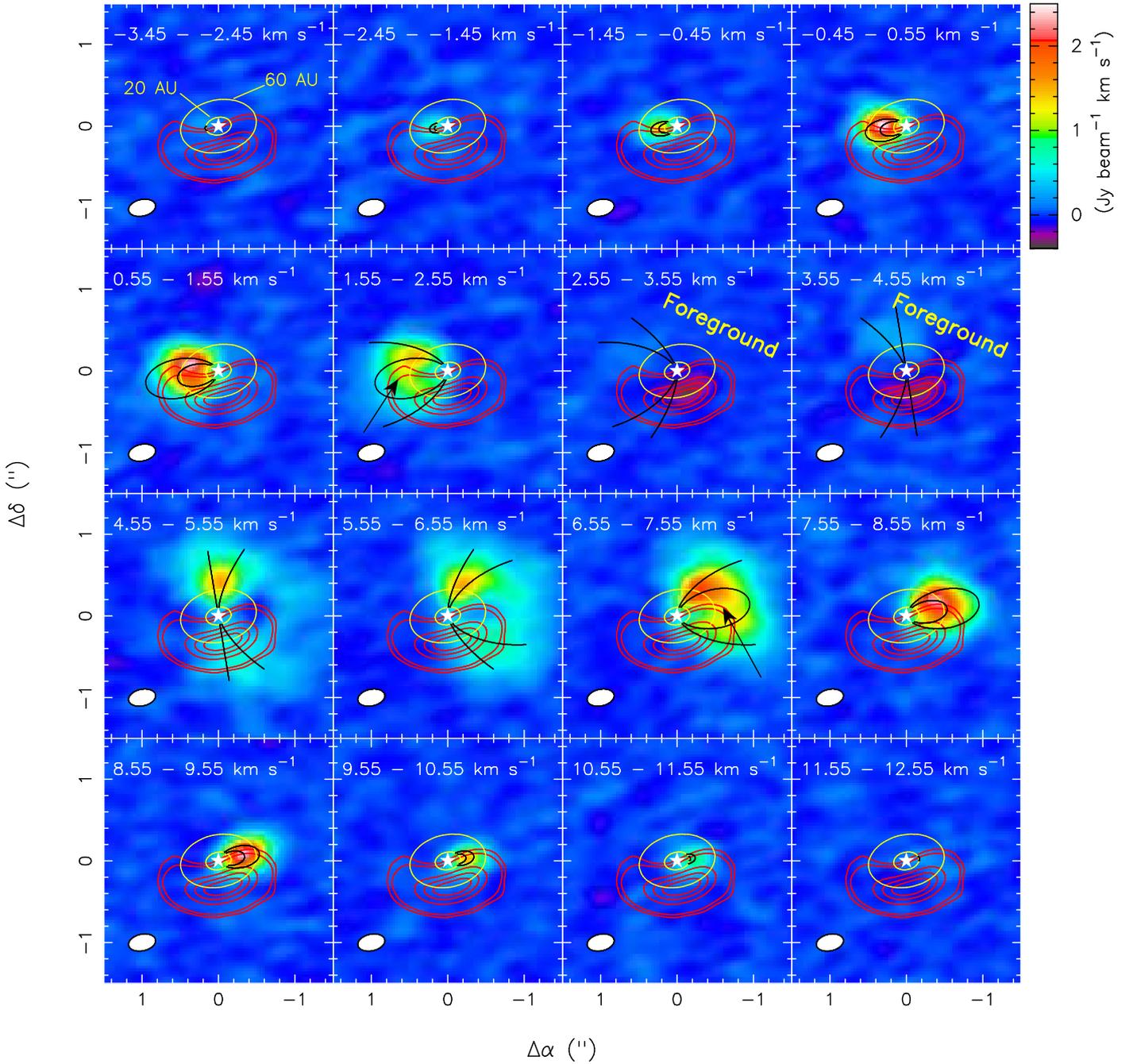}
\caption{$^{12}$CO $J=6-5$ channel map around ${\rm v}_{\rm source}=4.55$ km s$^{-1}$. Channels are binned to 1 km s$^{-1}$. Black lines show isovelocity contours of the border velocity of each velocity bin assuming Keplerian rotation with parameters in Table \ref{tab:basic_irs48}. Channels between 2.55 and 4.55 km s$^{-1}$ are affected by foreground absorption. Black arrows indicate regions with clear deviations from the Keplerian pattern (see Section \ref{sec:result_chan}). Circles indicated radii of 20 and 60 AU. Red contours give the continuum ($5\sigma, 10\sigma, 100\sigma, 200\sigma, 300\sigma$ detections).} 
\label{fig:velo_chan}
\end{figure*}

Channel maps of $^{12}$CO $6-5$, binned to 1 km s$^{-1}$, are presented in Figure \ref{fig:velo_chan}. Overlayed on the spectrum are isovelocity contours for the velocity bin, derived in the same way as for Figure \ref{fig:int_imgdeproj}b (Section \ref{sec:result_int}). The blue part at ${\rm  v}_{\rm lsr} < {\rm v}_{\rm source}=4.55$ km s$^{-1}$ shows in the fastest channels with CO detected the expected Keplerian pattern,  derived using the parameters in Table \ref{tab:basic_irs48}. At slower channels (${\rm  v}_{\rm lsr}= 1.55  - 2.55$ km s$^{-1}$) some emission along the major axis of the disk at distances smaller than expected from the Keplerian pattern is found (indicated by black arrows in Figure \ref{fig:velo_chan}). The reason for this slow gas is not clear, but could be related to gas streaming inwards. In our data, this component is only clearly seen at distances $>$60 AU and it is thus uncertain if it is related to gas streaming towards the star as found by \cite{Casassus13} for the case of \object{HD 142527}. Between ${\rm  v}_{\rm lsr} = 2.55$ and $4.55$ km s$^{-1}$, the emission is completely absorbed by foreground. The red-shifted channels with respect to ${\rm  v}_{\rm source}$ show a Keplerian pattern for channels between ${\rm  v}_{\rm lsr} = 4.55 - 6.55$ km s$^{-1}$ and the fastest channels (${\rm  v}_{\rm lsr}>8.55$ km s$^{-1}$). In between, some emission along the major axis at smaller distances than expected from the Keplerian pattern is found, as for the blue shifted side. 

We conclude that the disk for higher velocities and thus for regions closer to the star, follows the expected Keplerian rotation. Deviations from the Keplerian rotation are found at velocities slower than 4 km s$^{-1}$ relative to ${\rm v}_{\rm source}$ at regions outside $\sim$60 AU.

\subsection{Total spectrum and C$^{17}$O} \label{sec:result_spec_c17o}

\begin{figure}[htb!]
\center
\includegraphics[width=0.95\hsize]{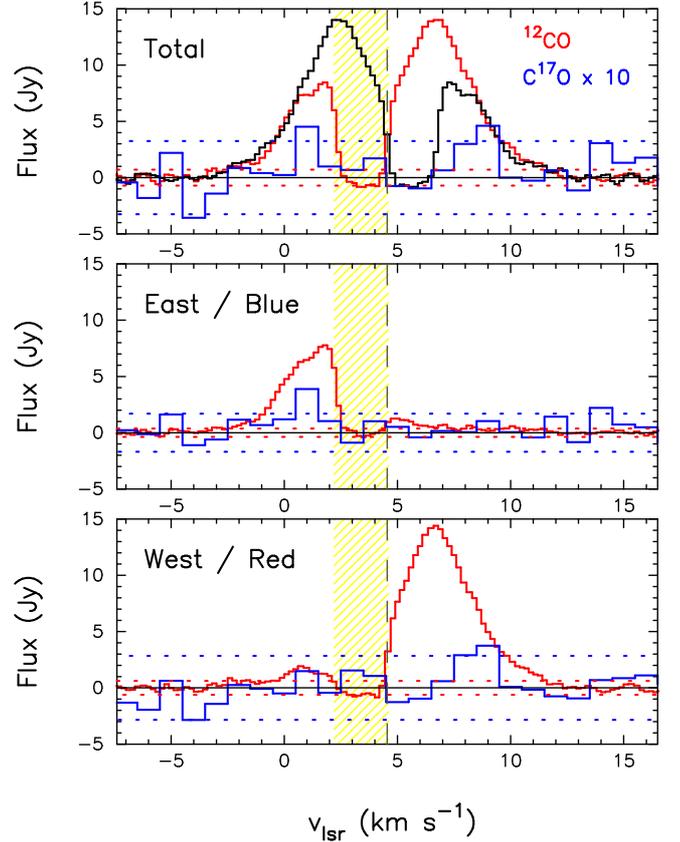}
\caption{Continuum subtracted spectra of $^{12}$CO $J=6-5$ (red lines) and $^{17}$CO $J=6-5$ (blue lines). The spectra are extracted from regions with $> 5 \sigma$ detections in $^{12}$CO. The yellow shaded region indicated velocities affected by foreground absorption. The vertical dashed line at 4.55 km s$^{-1}$ gives ${\rm v}_{\rm source}$. Dotted horizontal lines show the $3 \sigma$ detection level. \textit{Top panel:} Total spectrum. The spectrum of $^{12}$CO mirrored on ${\rm v}_{\rm source}$ is given in black line. \textit{Center panel:} Spectrum extracted from the east/blue part of the disk. \textit{Bottom panel:} Spectrum extracted from the west/red part of the disk.}
\label{fig:tot_spec}
\end{figure}

The total spectrum extracted from regions with $>$5$\sigma$ detections in $^{12}$CO is shown in Figure \ref{fig:tot_spec}, together with the spectrum mirrored at ${\rm v}_{\rm source}$. Original and mirrored spectrum overlay for velocities faster than $\sim$4 km s$^{-1}$ relative to v$_{\rm source}$, indicating a symmetric disk close to the protostar. Assuming Keplerian rotation, a velocity shift of 4 km s$^{-1}$ corresponds to radius of $r = \sin(i)^2 G M_* / {\rm v}^2 \sim 65$ AU. CO is detected out to velocity shifts of $\sim$7 km s$^{-1}$, corresponding to a radius of $\sim$20 AU. Between ${\rm v}_{\rm lsr}=0$ and 2 km s$^{-1}$, the blue shifted side is weaker than the red shifted side. This could be due to an additional foreground layer (Section \ref{sec:result_int}).

The emission of C$^{17}$O, observed in parallel with $^{12}$CO, is too weak to be detected in a single beam. When integrated over a larger region and binned to 1 km s$^{-1}$ velocity-resolution, it has been detected at a $\sim$4$\sigma$ level. Only channels with velocity of 3.5 and 4.5 km s$^{-1}$ relative to ${\rm v}_{\rm source}$ are detected. Only the channel at 4.5 km s$^{-1}$ is detected on both the blue and red side. The largest contribution to this channel is from radii of 60 to 110 AU, assuming Keplerian rotation. The lower panels in Figure \ref{fig:tot_spec} show spectra extracted from regions where $^{12}$CO is detected at a $>$5$\sigma$ level in the blue and red part of the spectrum. Indeed, the C$^{17}$O emission is also only detected on the corresponding side of the spectrum, providing some evidence that the C$^{17}$O detection is real.

We conclude from the total spectrum that the disk does not show clear signs of asymmetries inside 60 AU. Observations at higher angular resolution or better sensitivity may however reveal asymmetries also in this region. We further conclude that C$^{17}$O is detected at $>$60 AU, although at a very weak level.

\subsection{A cut through the major axis of the disk}  \label{sec:result_cut}

\begin{figure}[htb!]
\center
\includegraphics[width=0.95\hsize]{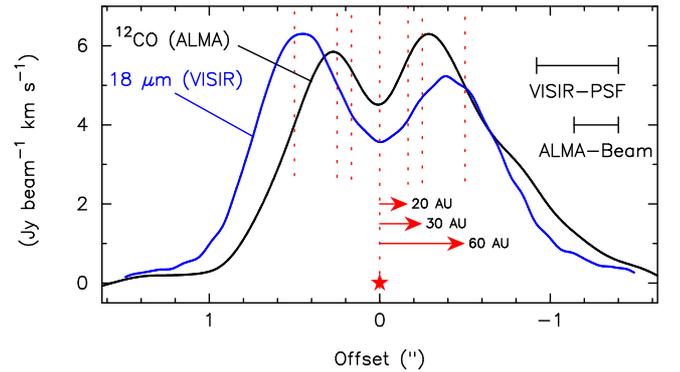}
\caption{Integrated intensity of $^{12}$CO $J=6-5$ (black line) extracted from a cut through the major axis of the disk ($y=0$ AU in Figure \ref{fig:int_imgdeproj}b). The 18.7 $\mu$m dust emission observed by VLT-VISIR (blue line) has been scaled to the $^{12}$CO peak.}
\label{fig:cutobs}
\end{figure}

The previous sections have shown that a cut through the major axis of the disk is least affected by foreground absorption. In Figure \ref{fig:cutobs}, the integrated intensity of $^{12}$CO along this cut is presented together with the 18.7 $\mu$m dust continuum emission observed with VLT-VISIR.

Both the $^{12}$CO integrated intensity and 18.7 $\mu$m dust continuum emission profiles are double-peaked, with the peaks located symmetrically with respect to the star. The strengths of the peaks are slightly asymmetric with the dust emission $\sim$15 \% stronger in the east, while the line emission is about $\sim$10 \% stronger in the west. The $^{12}$CO emission peaks at a distance of $\sim$35 AU to the star, while the dust emission peaks at $\sim$55 AU. At the positions where the dust emission peaks, no clear break or change in the $^{12}$CO emission can be seen. The depletion in the center of the $^{12}$CO emission has about the width of one beam. 

%%%%%%%%%%%%%%%%%%%%%%%%%%%%%%%%%%%%%%%%%%%%%%%%%%%%%%%%%%%%%%%%%%%%%%%%%%%%%%%%%%%%%%%%%%%%%%%%%
%
% Analysis
%
%%%%%%%%%%%%%%%%%%%%%%%%%%%%%%%%%%%%%%%%%%%%%%%%%%%%%%%%%%%%%%%%%%%%%%%%%%%%%%%%%%%%%%%%%%%%%%%%%
%%%%%%%%%%%%%%%%%%%%%%%%%%%%%%%%%%%%%%%%%%%%%%%%%%%%%%%%%%%%%%%%%%%%%%%%%%%%%%%%%%%%%%%%%%%%%%%%%

\section{Analysis} \label{sec:analysis}

In order to constrain changes of the gas surface density profile in particular the drop of the gas surface density inside the dust cavity and the inner gas hole, we use the combined physical-chemical model DALI (Dust And LInes) by \cite{Bruderer13}, based on \cite{Bruderer12}. The models solve for the dust radiative transfer, the chemical abundance, the molecular excitation and the thermal balance to obtain the gas temperature in a self-consistent way. Such a model is also needed to determine the abundance of CO inside and outside the dust cavity since it is not a priori obvious that CO can survive in the dust-free gas exposed to intense UV radiation from the star. The CO abundance distribution in the disk, together with the gas temperature distribution, determines the intensity and profile of the CO line emission. Spectral image cubes of the line emission are derived from the models and compared to the observations. 

The main focus of our modeling are the regions inside the 60 AU radius dust cavity. As discussed in Section \ref{sec:results}, this region does not show strong signs of asymmetries and we will restrict our analysis to 2d axisymmetric models. Fully 3d models might ultimately be needed to explain e.g. the structure of the outer disk, but they are computationally very demanding. Also, a more detailed analysis will require the additional constraints from optically thin isotopologue emission (\citealt{Bruderer13}) which are not yet available.

In this section, we will first discuss the adopted disk structure for our modeling, and then show the physical and chemical structure derived from the models and a comparison of the line emission to the observations.

\subsection{Disk structure} \label{sec:analysis_diskstruct}

\begin{figure}[htb!]
\center
\includegraphics[width=0.95\hsize]{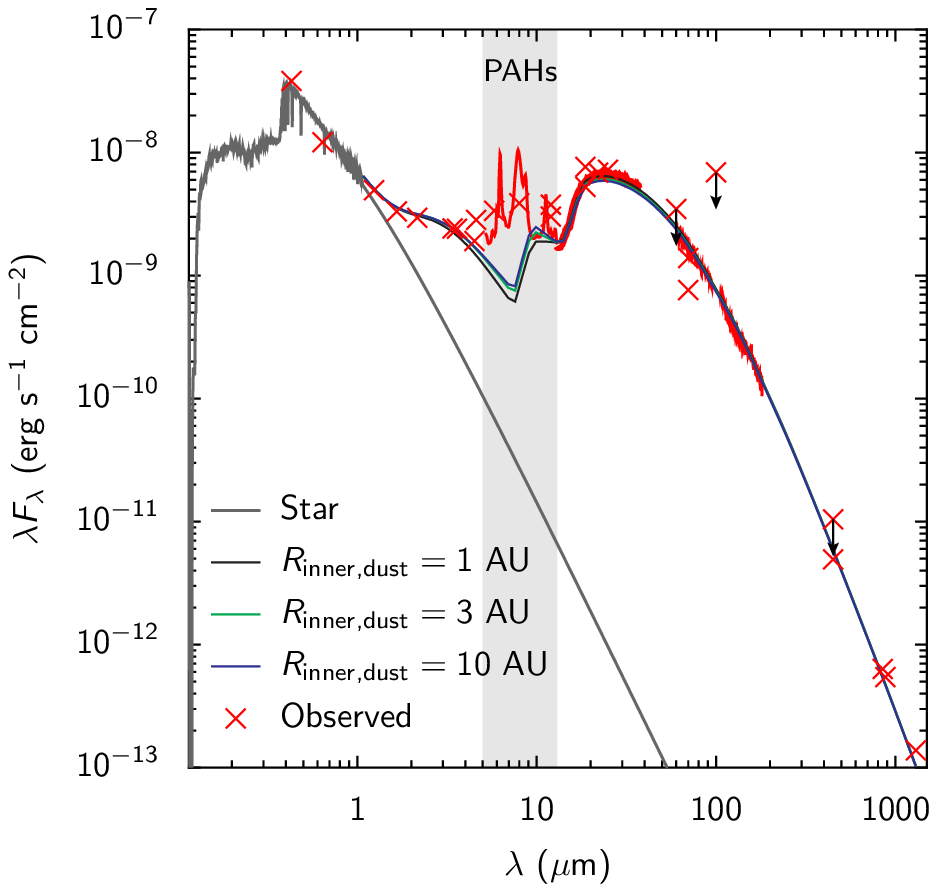}
\caption{Observed, dereddened SED of IRS 48 (red crosses/lines) compared to models with different extent (outer radius) of the warm dust inner disk ($R_{\rm inner,dust}=1, 3$ or 10 AU in black, green and blue lines). The stellar SED is shown by a grey line. Wavelengths with PAH emission not used in the SED fitting are indicated by a grey shaded region.}
\label{fig:sed}
\end{figure}

The dust structure of IRS 48 is derived from the SED and 18.7 $\mu$m VLT-VISIR dust continuum image by \citet{Geers07}. The SED is given in Figure \ref{fig:sed} (Table \ref{tab:fluxes}). It has been dereddened using a \citet{Weingartner01} extinction curve for $R_V$ = 5.5 assuming $A_V$ = 11.5 (\citealt{Brown12}). The stellar SED is derived from the \citet{Pickles98}-library spectrum for an A0 star with bolometric luminosity 14.3 L$_\odot$. At wavelengths shorter than 1150 \AA, this spectrum is extended by FUSE observations towards a star of the same spectral type, scaled for the bolometric luminosity. The dust SED shows excess emission over the stellar emission at near infrared wavelengths (1-5 $\mu$m), indicative of warm dust close to the star ($<$1 AU). The spatial location of this warm dust is not known, but is here attributed to the optically thick inner wall of a dusty inner disk. Other SED fits of IRS 48 (\citealt{Maaskant13}) have assigned the near infrared excess to a dust halo around the star. A peculiar feature of the SED are the strong PAH features at 5-12 $\mu$m, discussed by \citealt{Geers07}. The mid-infrared excess has been spatially resolved by VLT-VISIR at 18.7 $\mu$m and comes from a ring located at a radial distance of $\sim$60 AU to the star. The (sub)millimeter continuum is strongly lopsided, but remains optically thin at 450 $\mu$m (\citealt{vdMarel13}). The peak emission of the submillimeter continuum is located slightly outside 60 AU (Figure \ref{fig:int_imgdeproj}, \citealt{vdMarel13}) and in the following we will assume that both the mid-infrared and (sub)millimeter continuum emission come from an outer disk located at a radius $>$60 AU. The FIR continuum (60-180 $\mu$m), observed by Herschel (\citealt{Fedele13a}), is not spatially resolved and it is unknown whether it is ring-like or lopsided.

\begin{table}[tbh]
\caption{Observed continuum fluxes}
\label{tab:fluxes}
\centering
\begin{tabular}{llrl}
\hline \hline
Instrument & Wavelength & Flux & Reference\\
 & ($\mu$m) & (Jy) \\
\hline
\multicolumn{4}{c}{Photometry} \\ 
\hline
               NOMAD &                 0.43 & 1.60(-5) & \cite{Zacharias05} \\
               NOMAD &                 0.64 & 3.26(-4) & \cite{Zacharias05} \\
               2MASS &                 1.24 & 9.46(-2) & \cite{Cutri03} \\
               2MASS &                 1.66 & 3.05(-1) & \cite{Cutri03} \\
               2MASS &                 2.16 & 6.18(-1) & \cite{Cutri03} \\
                WISE &                 3.4  & 1.29     & \cite{Wright10} \\ 
                IRAC &                 3.6  & 1.41     & \cite{vanKempen09} \\
                IRAC &                 4.5  & 1.60     & \cite{vanKempen09} \\
                WISE &                 4.6  & 2.45     & \cite{Wright10} \\ 
                IRAC &                 5.8  & 4.06     & \cite{vanKempen09} \\
                IRAC &                 8.0  & 6.00     & \cite{vanKempen09} \\
                IRAS &                12.0  & 7.81     & \cite{Helou88} \\
                WISE &                12.0  & 6.26     & \cite{Wright10} \\ 
               AKARI &                18.7  & 21.0     & \cite{Yamamura10} \\
               VISIR &                18.7  & 31.8     & \cite{Geers07} \\
                WISE &                22.0  & 36.5     & \cite{Wright10} \\ 
                IRAS &                25.0  & 46.5     & \cite{Helou88} \\
                IRAS &                60.0  & $<$65.5  & \cite{Helou88} \\                
                MIPS &                70.0  & 17.0     & \cite{vanKempen09} \\
                PACS &                70.0  & 31.0     & \cite{Fedele13a} \\
                IRAS &               100.0  & $<$225.0 & \cite{Helou88} \\                
                ALMA &               450.0  & 9.50(-1) & \cite{vdMarel13} \\
               SCUBA &               450.0  & $<$1.57  & \cite{Andrews07} \\                              
               SCUBA &               850.0  & 1.80(-1) & \cite{Andrews07} \\               
                 SMA &               880.0  & 1.60(-1) & \cite{Andrews07} \\               
                 SMA &              1300.0  & 6.00(-2) & \cite{Andrews07} \\               
\hline                 
\multicolumn{4}{c}{Spectroscopy} \\ 
\hline                          
IRS                  & \multicolumn{2}{l}{5.9 $-$ 36.89            } & \cite{McClure10} \\
PACS                 & \multicolumn{2}{l}{60.0 $-$ 73.3 (B1)       } & \cite{Fedele13a} \\
                     & \multicolumn{2}{l}{69.8 $-$ 95.0 (B1)       } & \cite{Fedele13a} \\
                     & \multicolumn{2}{l}{105.0 $-$ 146.6 (R1)     } & \cite{Fedele13a} \\
                     & \multicolumn{2}{l}{139.6 $-$ 180.9 (R2)     } & \cite{Fedele13a} \\
\hline                     
\end{tabular}
\tablefoot{a(b) means $a \times 10^{b}$}
\end{table}

For our modeling, we adopt the physical structure suggested by \citet{Andrews11}, as implemented by \citet{Bruderer13}. The surface density profile is assumed to be a power-law and the vertical structure follows a Gaussian distribution (vertically isothermal structure). Given that our data do not constrain the vertical structure and disk models assuming hydrostatic equilibrium overestimate the scale-height (\citealt{Thi13}), we feel that such a simple approach is warranted. The outer radius of the disk is not the focus of our work, and we here employ a sharp cut-off in the surface density profile rather than an exponential decrease as in \citet{Andrews11} and \citet{Bruderer13}.

\begin{figure}[htb!]
\center
\includegraphics[width=0.95\hsize]{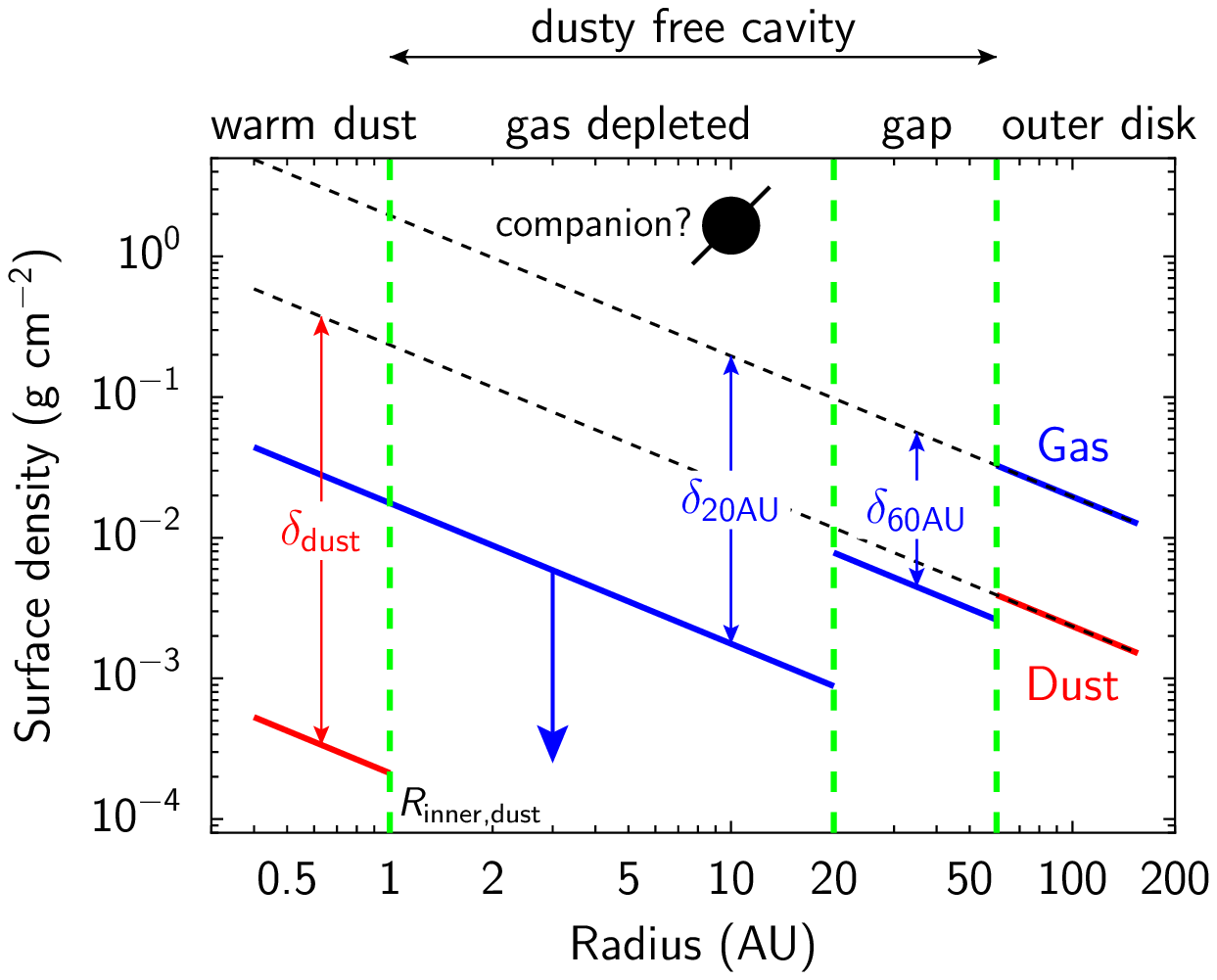}
\caption{Surface density profile of the representative model. Gas and dust surface density are given in blue and red lines, respectively. The black dashed lines show the gas/dust surface density if there were no cavities (Eq. \ref{eq:surfdens}). Green dashed lines indicate the outer radii of the inner warm disk, the gas depleted hole, and the dust free cavity (Table \ref{tab:params}) In the gas depleted hole, only upper limits on the gas surface density profile can be derived from the observations.}
\label{fig:radstruct}
\end{figure}

The parameters of the structure are shown in Table \ref{tab:params} and Figure \ref{fig:radstruct}. The surface density profile follows
\begin{equation} \label{eq:surfdens}
\Sigma(r) = \Sigma_{\rm60\, AU} \cdot \left(\frac{r}{\rm 60\, AU} \right)^{-\gamma} 
\end{equation}
and is defined by the power-law index $\gamma$ and a surface density $\Sigma_{\rm60\, AU}$ at 60 AU. The surface density profile of the outer disk at 60-160 AU is scaled by different factors within the warm dust inner disk inside $R_{\rm inner, dust}=1$ AU ($\delta_{\rm dust}$), the gas depleted hole inside 20 AU and the gap between 20-60 AU ($\delta_{\rm 20\, AU}$ and $\delta_{\rm 60\, AU}$). In the radial direction, these regions are defined by the dust sublimation temperature (0.4 AU), the line wings of the $^{12}$CO emission (20 AU), and the size of the dust cavity seen in the 18.7 $\mu$m dust continuum images and the (sub)millimeter continuum (60 AU). The critical parameter to be studied in this work are the drop of the gas surface density profile at 20 AU ($\delta_{\rm 20\, AU}$) and at 60 AU ($\delta_{\rm 60\, AU}$).

The vertical distribution is defined by a scale-height angle $h_{\rm 60\, AU}$ at 60 AU, such that the scale-height angle follows $h(r) = h_{\rm 60\, AU} \cdot (r/{60\,{\rm AU}})^\psi$. We use the same dust opacities as \citet{Andrews11} consisting of a small ($0.005-1$ $\mu$m) and large ($0.005$ $\mu$m - 1 mm) population of dust with a (mass) fraction $f$ in large grains and the scale-height of the large grains reduced by a factor $\chi$. We fix $f$ and $\chi$ to the values adopted by \citet{Andrews11} for other transition disks. A considerably lower value of $f$ is unlikely given the evidence found for grain-growth towards IRS 48 (\citealt{vdMarel13}). Since most of the UV opacity, which controls the gas temperature, is due to the small grain population, the scale height factor $\chi$ of the larger grains does not affect the line emission much.

\begin{table}[tbh]
\caption{Parameters of the representative model. The parameters are explained in Section \ref{sec:analysis_diskstruct} and shown in Figure \ref{fig:radstruct}.}
\label{tab:params}
\centering
\begin{tabular}{lll}
\hline\hline
Parameter &  &\\
\hline
Surface density profile\tablefootmark{1}    &$\gamma$                 & 1.0              \\
                                            &$\Sigma_{\rm 60\, AU, gas}$   & $3.2 \times 10^{-2}$  g cm$^{-2}$ \\
                                            &$\Sigma_{\rm 60\, AU, dust}$  & $4.0 \times 10^{-3}$  g cm$^{-2}$ \\
\hline                                                                                 
Radial sizes               & Inner warm dust         & $0.4 - 1$ AU \\
                           & Gas depleted hole       & $0.4 - 20$ AU \\
                           & Dust free cavity        & $1 - 60$ AU   \\
                           & Outer disk              & $60 - 160$ AU \\
\hline
Scaling of $\Sigma(R)$     & Dust, 0.4-1 AU          & $\delta_{\rm dust}=9 \times 10^{-4}$ \\
                           & Gas, 0.4-20 AU          & $\delta_{\rm 20\, AU} < 9 \times 10^{-3}$ \\          
                           & Gas, 20-60 AU           & $\delta_{\rm 60\, AU} = 8 \times 10^{-2}$ \\
\hline
Vertical structure         & $h_{\rm 60\, AU}$       & 0.14 radians\\
                           & $\psi$                  & 0.22 \\
\hline
Dust settling              & $f$                     & 0.85            \\
                           &$\chi$                   & 0.2             \\
\hline
\end{tabular}
\tablefoot{\tablefoottext{1}{See Eq. \ref{eq:surfdens}.}}
\end{table}

In order to constrain the dust structure of IRS 48, we have explored the dependence of the model SED on various parameters (e.g. surface density profile, scale-height function). Here, we disregard wavelengths affected by PAH emission (5-12 $\mu$m) and do not attempt to fit the PAH abundance. For the combined physical-chemical model we adopt a PAH abundance of $\sim$10 \% of the ISM abundance (5 \% PAH-to-dust mass ratio, \citealt{Draine07}) derived by comparing the feature-to-continuum ratio provided by Figure 7 in \cite{Geers06} to the observed strength of the features.

Exploring the model parameters, we find that reproducing the near infrared excess requires an optically thick dusty inner disk. The amount of excess is determined by the surface of optically thick hot dust at the inner edge of this inner disk. This optically thick layer however shields the wall of the outer disk at 60 AU and decreases the mid-infrared emission. In order to simultaneously fit the near- and mid-infrared emission, the scale-height of the outer disk needs to be high enough (scale-height angle $h_{\rm 60\, AU}=0.14$ radians) to intercept direct stellar irradiation. The disk mass is mostly constrained through the optically thin submillimeter emission. Assuming that the submillimeter emitting dust was originally spread out over the entire disk, the resulting total dust mass is $1.6 \times 10^{-5}$ M$_\odot$ ($\kappa_{450\, \mu{\rm m}}=7.1$ cm$^2$ g$^{-1}$). The inner disk mass is with a dust mass of $8 \times 10^{-11}$ M$_\odot$ much less massive. The dust mass of the inner disk is not well determined because the near infrared emission traces the optically thick inner wall of the inner disk. The 18.7 $\mu$m emission of the inner disk is not optically thick and a larger inner disk with a radius of 10 AU and a mass of $1 \times 10^{-9}$ M$_\odot$ would overproduce the 18.7 $\mu$m emission at the stellar position. The submillimeter emission, mapped by ALMA, is not axisymmetric. However, since the dust temperature in the mid-plane remains $>$50 K out to radii of $\sim$120 AU and the submillimeter emission is optically thin, the submillimeter opacities are not important in determining the dust temperature. Thus, our axisymmetric approach does not alter the derived dust mass much.

\begin{figure*}[htb!]
\center
\includegraphics[width=1.0\hsize]{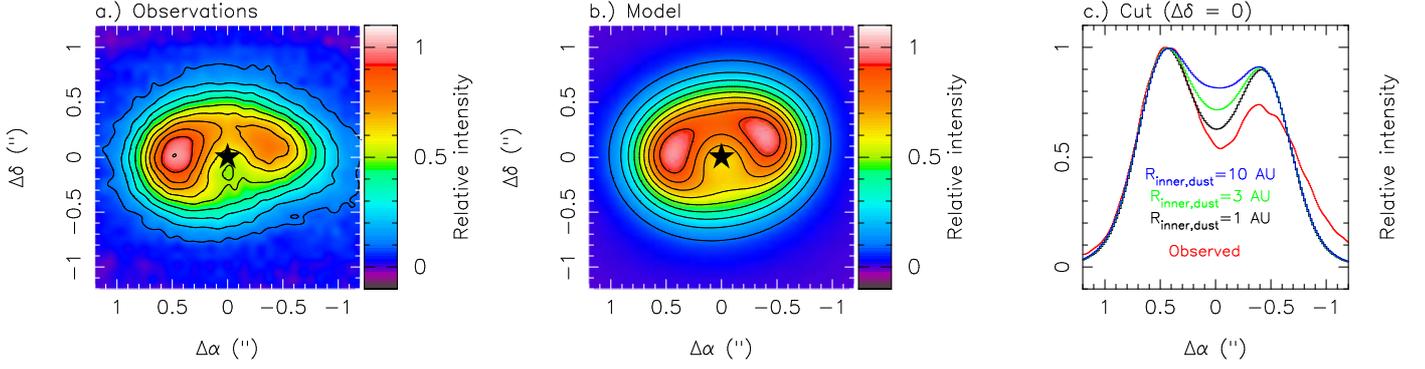}
\caption{\textit{Left/center panel:} Observed and modeled VLT-VISIR 18.7 $\mu$m image. Contour lines show $10, 20, \ldots$ \% of the peak emission. \textit{Right panel:} Cut through the observed and modeled images ($\Delta \delta = 0$). Red line: observed intensity. Black, green, and blue lines: modeled intensities for different outer radii of the warm dust inner disk ($R_{\rm inner,dust}=1, 3$ or 10 AU).} \label{fig:visir_irs48}
\end{figure*}

Figure \ref{fig:visir_irs48} compares the VLT-VISIR 18.7 $\mu$m observations to model images. The observations show a ring-like feature with a depletion at the stellar position and gradients both in east-west and north-south direction. The gradient in north-south direction is also seen in the models and is the result of the disk's scale-height in combination with the inclination. The dust emission at 18.7 $\mu$m in the outer disk is optically thick and traces the surface of the disk. Thus, while both near- and far-side of the northern part of the ring is visible to us, only the near-side is visible to us in the south (\citealt{Dullemond10}) resulting in a north-south asymmetry. Since the inner dusty disk at radii $<$10 AU is warm enough to emit at 18.7 $\mu$m, some constraints on the size of this inner dusty disk can be obtained from the image. If the inner disk has an size of 3 or 10 AU, the emission at the position of the star is too strong. Hence we choose $R_{\rm inner,dust}=1$ AU. Varying the radius of the inner disk has only a small effect on the SED (Figure \ref{fig:sed}), since the total emission at this wavelength is dominated by the ring at $\sim$60 AU. The reason for the east-west asymmetry is unclear, but could be due to azimuthal changes of the scale-height at the inner disk (\citealt{Espaillat11,Flaherty12}).

\subsection{Physical-chemical modeling} \label{sec:analysis_tcmodel}

Based on the dust structure derived in the previous section, we run the physical-chemical models for different gas surface density profiles. In the following sections we refer to the representative model as that with the gas surface density following the values given in Figure \ref{fig:radstruct} and Table \ref{tab:params}.

\begin{figure*}[htb!]
\center
\sidecaption
\includegraphics[width=0.7\hsize]{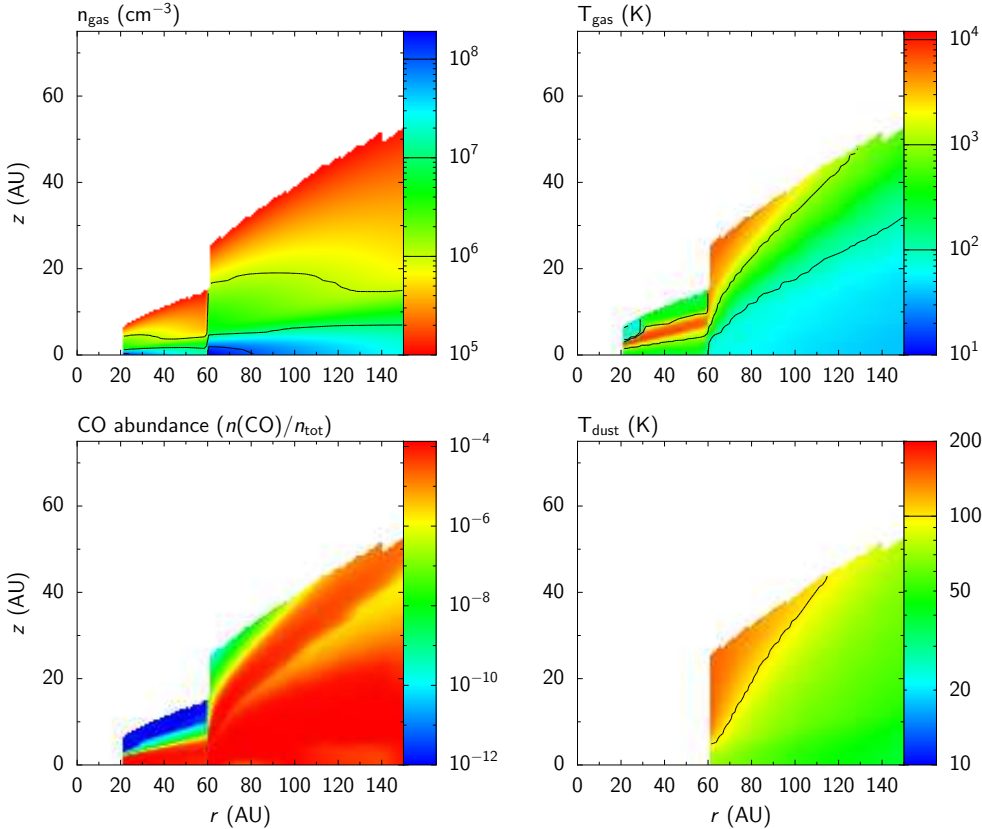}
\caption{Density structure, gas- and dust-temperature, and CO fractional abundance of the representative model (Table \ref{tab:params}). Only regions with gas density larger than $10^5$ cm$^{-3}$ are shown. The dust temperature is not defined in the dust free cavity. The z-axis is stretched by a factor of two compared to the r-axis.}
\label{fig:model_gasmass}
\end{figure*}

Figure \ref{fig:model_gasmass} shows the gas density, CO fractional abundance, gas-temperature ($T_{\rm gas}$) and dust-temperature ($T_{\rm dust}$) of the representative model. In this section we focus on the main features of the IRS 48 model, general trends of transition disk models are discussed in \citet{Bruderer13}.

Inside the dust cavity at radii $<$60 AU, the gas density in the mid-plane reaches $2 \times 10^8$ cm$^{-3}$. This is far higher than the critical density of CO $6-5$ of $\sim$10$^5$ cm$^{-3}$ and the observed molecular emission of CO $J=6-5$ is close to local thermal equilibrium (LTE). Temperatures in the dust cavity are several 1000 K in the upper atmosphere but still above 100 K in the mid-plane and thus high enough to excite CO $6-5$ with an upper level energy of 115 K. CO is photodissociated in the the upper atmosphere by the intense FUV radiation of the star. Below a narrow C$^+$/CO transition, carbon is fully bound in CO closer to the mid-plane.

In the outer disk, the gas- and dust temperature close to the mid-plane are well coupled and of order 50 K. At larger heights, the gas temperature decouples from the dust temperature and can reach temperatures of several 1000 K (\citealt{Bruderer12}). This hot layer leads to a ``warm finger'' of CO, where CO formation is initiated by the reaction of C$^+$ with vibrationally excited H$_2$ (e.g. \citealt{Jonkheid07}). 

\subsection{Comparison to observations}  \label{sec:analysis_compobs}

To compare model results with observations and to determine the drop in gas surface density at 20 AU ($\delta_{\rm 20 \, AU}$) and at 60 AU ($\delta_{\rm 60 \, AU}$), Figure \ref{fig:surfcomp} shows the $^{12}$CO $J=6-5$ integrated intensity cuts along the major axis of the disk for various trial gas distributions. The major parameters that are varied are $\Sigma_{\rm 60\, AU, gas}$, $\delta_{\rm 60 \, AU}$ and $\delta_{\rm 20 \, AU}$. The integrated intensity cut is chosen along the major axis, because this direction is least affected by foreground absorption (Section \ref{sec:result_cut}). In addition to the integrated intensity cut, the total spectrum is presented. The synthetic spectral image cubes from the models are convolved to the ALMA beam. To verify this approach, the result of the representative model convolved to the ALMA beam is compared to the result of simulated ALMA observations (Online-Figure \ref{fig:simul}). The ALMA observations have been simulated using the CASA software by converting the model image to ${\rm (u,v)}$-data according to the observed ${\rm (u,v)}$-coverage and then reduced in the same way as the observations. Due to the good ${\rm (u,v)}$-coverage of our observations, the two approaches do not differ measurably in the intensity cut and the total spectrum at velocities $\geq 4$ km s$^{-1}$ from the source velocity, which is the focus of this work.

The line center opacity of $^{12}$CO is of order 20 in the center of the dust free cavity (40 AU) and of order 100 in the outer disk at 60 AU for the representative model (Table \ref{tab:params}). It is thus clear that the current observations cannot directly constrain the surface density profile (\citealt{Bruderer13}). However, due to temperature changes in the vertical structure and the fact that parts of the lines remain optically thin due to the Keplerian rotation, optically thick lines still show some dependence on the gas surface density profile and in particular on changes like drops inside the dust free cavity. The absolute scaling of the derived surface density profile can be checked with the optically thin emission of C$^{17}$O.  We do not expect that derived surface density profiles are more accurate than a factor of a few. To test our proposed surface density profile with future observations, we will discuss the derived emission of the optically thin isotopologues in Section \ref{sec:coisotop}.

The disk exhibits asymmetries at radii larger than 60 AU (Section \ref{sec:results}). Most of the emission from these regions of the disk emerge at low velocities in the spectrum ($<$4 km s$^{-1}$ relative to v$_{\rm source}$). We ignore these asymmetries here and focus the model-observation comparison to the inner part of the disk and higher velocities.

\subsubsection{The outer disk ($\Sigma_{\rm 60 AU,gas}$)}  \label{sec:analysis_outerdisk}

Figure \ref{fig:surfcomp}a shows the $^{12}$CO integrated intensity cut and total spectrum of the representative model (Figure \ref{fig:radstruct} and Table \ref{tab:params}) and two models with a factor of 10 increased or decreased gas surface density profile in the outer disk (varied $\Sigma_{\rm 60 AU,gas}$). The gas mass is likely dominated by the outer disk, and the outer disk surface density profile thus constrains the total gas mass. The representative model with $\Sigma_{\rm 60 AU,gas}=3.2 \times 10^{-2}$ g cm$^{-2}$ (red lines in Figure \ref{fig:surfcomp}) yields an integrated intensity along the major axis between the observed east/west integrated intensity. The total spectrum derived from this model also agrees relatively well to the observed spectrum at high velocities ($\geq$4 km s$^{-1}$ from the source velocity), corresponding to the inner disk. Slower velocity channels, mostly corresponding to the outer disk, show some deviations. The model emission is too strong at $0.5-2$ km s$^{-1}$, possibly due to an additional foreground layer and at $5 - 8$ km s$^{-1}$ due to azimuthal variations of the outer disk (Section \ref{sec:results}).

Models with a factor of 10 higher column density overproduce both the integrated intensity cut and the total spectrum by more than a factor of two. The model with a factor of 10 lower column density on the other hand, underproduces the integrated intensity at offsets between 0.5'' and 0.8'' (60-100 AU). While the total spectrum of this model better reproduces the wings at around 0.5 and 8.5 km s$^{-1}$, it underproduces the line center. 

\begin{figure}[htb!]
\center
\includegraphics[width=0.78\hsize]{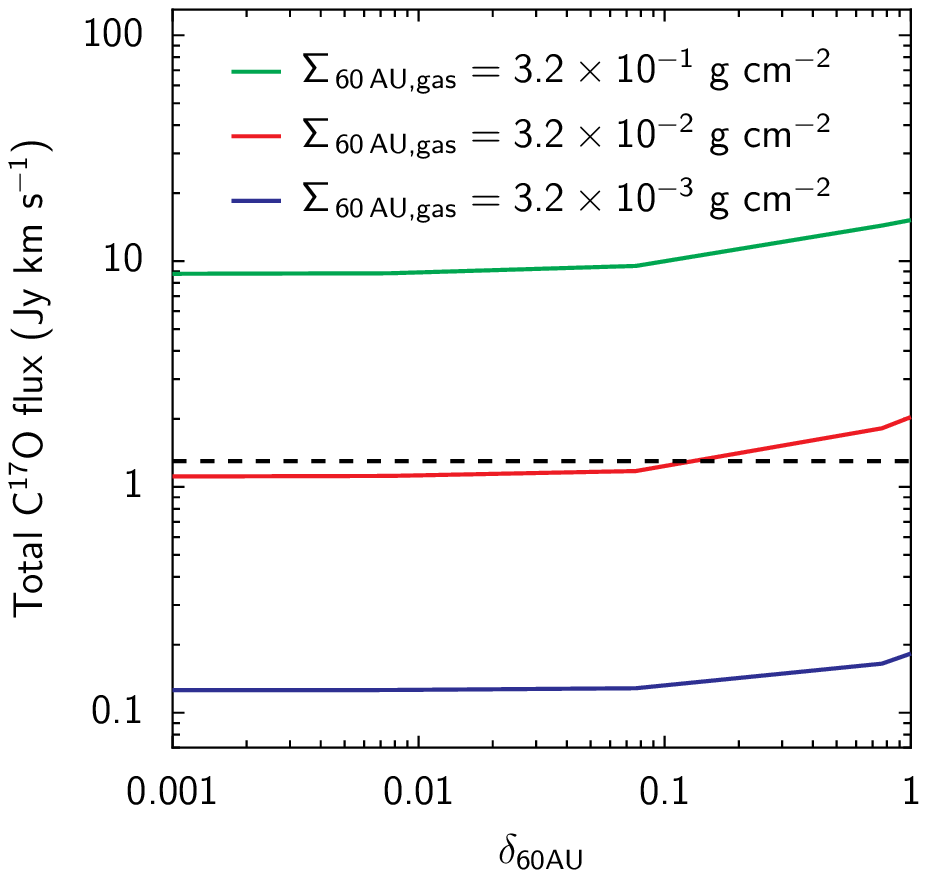}
\caption{Total flux of C$^{17}$O for models with $\Sigma_{\rm 60\, AU, gas}=3.2 \times 10^{-1}, 3.2 \times 10^{-2}$ and $3.2 \times 10^{-3}$ g cm$^{-2}$ as a function of the gas surface density at 20-60 AU ($\delta_{\rm 60\, AU}$). The dashed vertical line shows the observed total flux. The red line indicates the representative model.}
\label{fig:c17o_flux}
\end{figure}

The C$^{17}$O total flux yields another, independent constraint on the total mass of the disk and thus the surface density profile of the outer disk. Figure \ref{fig:c17o_flux} compares the modeled C$^{17}$O total flux of the three models discussed before with the observed 1.3 Jy km s$^{-1}$. We show the C$^{17}$O flux for three values of $\Sigma_{\rm 60\, AU, gas}$, varying the amount of gas inside 60 AU ($\delta_{\rm 60\, AU}$). The total gas mass is dominated by the gas mass outside 60 AU for $\delta_{\rm 60\, AU} \lesssim 0.1$ and rises by only $\sim$50 \% for $\delta_{\rm 60\, AU}=0.1-1$. It is seen that the representative model reproduces the observed flux to within $\sim$20 \%, while the models with increased/reduced surface density over/underproduce the emission. As C$^{17}$O remains optically thin and thus directly traces the gas mass, its emission scales with the gas mass.

We conclude that the representative model reproduces relatively well the $^{12}$CO integrated intensity cut and total spectrum together with the optically thin C$^{17}$O emission. Some deviations in the $^{12}$CO total spectrum are due to azimuthal asymmetries of the disk, not included in our model. The gas mass\footnote{This gas mass is much lower than that derived by \citet{vdMarel13}. In that work, we have mistakenly multiplied the number of H$_2$ molecules derived from the C$^{17}$O observations with the mass of C$^{17}$O instead of the H$_2$ mass.} of the representative model is $1.4 \times 10^{-4}$ M$_\odot$. This gas mass is low, only $\sim$15 \% of the mass of Jupiter. Combined with the dust mass of $1.6 \times 10^{-5}$ M$_\odot$ (Section \ref{sec:analysis_diskstruct}), this gives a gas-to-dust mass ratio of about 10. Both gas and dust mass have an uncertainty at a level of a factor of a few.

\begin{figure*}[htb!]
\center
\includegraphics[width=1.0\hsize]{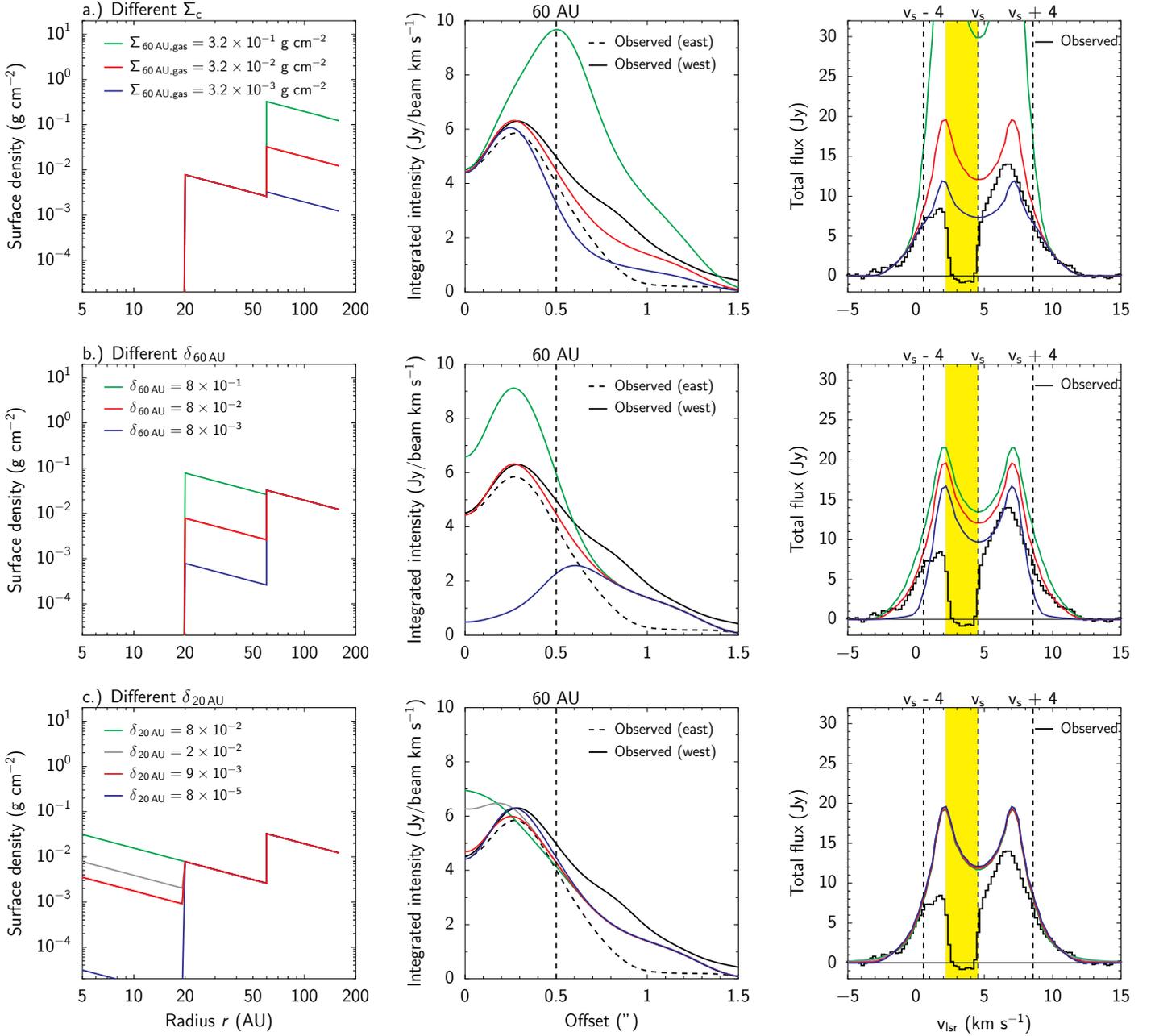}
\caption{Comparison of modeled and observed $^{12}$CO $J=6-5$ intensity profiles and spectra. \textit{Left panels:} Adopted surface density profiles with variations of the representative model. The red line is the representative model. \textit{Center panel:} Intensity profile extracted along the major axis of the disk. The dashed vertical line indicates the radius of the dust free cavity. \textit{Right panel:} Total spectrum. The yellow region indicates regions affected by the foreground. Vertical dashed lines show the source velocity ${\rm v}_{\rm s}=4.55$ km s$^{-1}$ and velocities ${\rm v}_{\rm s} \pm 4$ km s$^{-1}$ representing velocity ranges corresponding to the inner disk (radii $\lesssim 60$ AU) \textit{a)} Different surface density profiles in the outer disk ($>$60 AU). \textit{b)} Different surface density profiles inside the dust gap ($20 - 60$ AU). \textit{c)} Different surface density profiles inside 20 AU. } \label{fig:surfcomp}
\end{figure*}

\subsubsection{Inside the gap ($\delta_{\rm 60\, AU}$)}  \label{sec:analysis_drop60au}

Figure \ref{fig:surfcomp}b shows the model-observation comparison for models with the surface density profile between 20 and 60 AU varied ($\delta_{\rm 60\, AU}$ changed). The representative model is compared to a model with a factor of 10 more/less gas in this region. While the representative model reproduces the integrated intensity cut at offsets $<$0.5$''$ well, the models with a factor of 10 higher/lower surface density at $20-60$ AU over/underproduce the integrated intensity. The integrated intensity cut depends on the surface density inside 60 AU out to a radius of $\sim$90 AU (0.75''), due to the width of the beam. The total spectrum of the model with a factor of 10 higher surface density is too broad in the wings and the spectrum with a factor of 10 lower surface density is too narrow. 

We conclude that a drop in surface density at 60 AU by about a factor of $\sim$12 is required to fit the integrated intensity profile and reproduce the line width of the total spectrum. 

\subsubsection{The inner 20 AU ($\delta_{\rm 20\, AU}$)}  \label{sec:analysis_drop20au}

The innermost 20 AU have been suggested by \cite{vdMarel13} to be gas depleted, based on the drop in integrated intensity and the $^{12}$CO line wings. They speculated that this drop is related to the presence of an undetected planet or substellar companion. Since the size of this hole is comparable to the beam size (Figures \ref{fig:int_imgdeproj}, \ref{fig:velo_chan}, and \ref{fig:cutobs}) and a beam centered on the star will receive some of the much stronger emission from outside 20 AU, the current data can only set limits on the gas mass inside 20 AU.

Figure \ref{fig:surfcomp}c shows the representative model with different drops of the gas surface density profile inside 20 AU (varied $\delta_{\rm 20\, AU}$). In absence of any drop of the gas surface density profile at 20 AU, the dip in integrated intensity at the stellar position is not seen. For this amount of gas, the region outside 20 AU is slightly shielded from stellar radiation and the integrated intensity out to an offset of 0.5'' decreases. Only models with a drop of surface density $\delta_{\rm 20\, AU} < 2 \times 10^{-2}$ (factor of $>$50) yield a dip in integrated intensity at the stellar position. A model with a drop $\delta_{\rm 20\, AU} < 9 \times 10^{-3}$ (factor of $>$110) reproduces the observed dip at the center position. This model and a model with a two orders of magnitude larger drop show very similar integrated intensity cuts and total spectra. Thus, no additional constraints on the gas inside 20 AU can be drawn from the total spectrum. 

We conclude that the current data allows us to constrain a drop of a factor of $\gtrsim$110 in gas surface density profile at 20 AU. This corresponds to a factor of $\gtrsim$10 drop in surface density profile at 20 AU compared to the surface density at 20 to 60 AU. Better constraints will require higher angular resolution observations. 

\subsection{The slope $\gamma$ of the surface density profile} \label{sec:gammasurf}

To determine to what extent the assumed slope of the gas and dust surface density profile affects the derived properties, we have also run models with $\gamma=0$ and $\gamma=2$. The SED and 18.7 $\mu$m images can be equally well fitted for these different values of $\gamma$. Starting from the refitted dust structures, the gas surface density profile has been determined (Online-Figure \ref{fig:gamma}).

In order to fit the C$^{17}$O emission, the disk mass of the model with $\gamma=0$ and $\gamma=2$ needs to be adjusted to the  same value as for the model with $\gamma=1$. For the model with $\gamma=2$ and $\gamma=0$, we constrain $\delta_{\rm 60\, AU}=3 \times 10^{-2}$ and $\delta_{\rm 60\, AU}=0.25$, respectively. For $\gamma=0$, the emission at the stellar position is slightly underproduced, while at a distance of $\sim$ $0.2''-0.3''$ it is already overproduced. We conclude that the results for the gas mass and the presence of a drop at 60 AU are robust, but that the slope of the surface density profile cannot be well determined by the current data set.

The direct measurement of the surface density slope using CO and isotopologues is challenging, since different slopes only lead to a factor $\sim$2 difference in surface density over the range of radii. In Figure \ref{fig:iso}, the integrated intensity cut of CO and isotopologues is presented for different values of $\gamma$ and different beam sizes. The figure shows that isotopologue lines indeed directly trace the gas surface density inside the dust free 60 AU cavity. In the outer disk, however, shielding effects by dust become important leading to a slightly different CO abundance and it is difficult to trace the column density to the required accuracy to determine $\gamma$.

\subsection{How well is the 20 AU radius determined?} \label{sec:20audeterminded}

How well can the inner radius of the gas disk at 20 AU be determined? To answer this question, we run models with an inner radius of 15, 20, or 25 AU. Models with the surface density profile of the representative and with the surface density profile adjusted such that the emission at the stellar position is approximately reproduced are calculated. The model-observation comparison of these models is shown in Online-Figure \ref{fig:surfcomp2} in the same way as Figure \ref{fig:surfcomp}.

We find that models with an inner cavity of 15 or 25 AU instead of 20 AU are unable to reproduce the observed integrated intensity cuts. Models with the gas surface density profile from the representative model fail to reproduce the integrated intensity at the stellar position, while models with a surface density profile adjusted for the emission at the stellar position over or underproduce the emission at offsets of 0.2'' - 0.5''. We conclude that the drop of the surface density at 20 AU is constrained to better than $\sim$5 AU.

\subsection{The presence of an inner gas disk} \label{sec:inner_gasdisk}

From the fastest channels with $^{12}$CO detected, we concluded that a gas hole with radius 20 AU surrounds the star. How much gas could be hidden inside this hole, located in a small gas disk close to the star, similar to the warm dust disk needed to reproduce the near infrared excess in the SED? A series of models with $\delta_{\rm 20\, AU}$ varied, but gas completely removed between $1-20$ AU or $10-20$ AU is run. 

We find that for an inner gas disk size of 10 AU, $\delta_{\rm 20\, AU} > 8 \times 10^{-3}$ overproduces the central peak, while the spectrum is not affected much (Online-Figure \ref{fig:surfcomp3}). For $\delta_{\rm 20\, AU} \sim 8 \times 10^{-3}$, the PAHs in the inner disk shield the disk at 20 AU decreasing the integrated intensity slightly (by $\sim$10 \%). For an inner gas disk size of 1 AU, only this effect of the shielding can be seen, and even $\delta_{\rm 20\, AU} = 1$ would not produce additional emission at the position of the star. The upper limits on the CO mass of an inner disk are $5 \times 10^{23}$ g for a 10 AU disk with $\delta_{\rm 20\, AU} = 8 \times 10^{-3}$ and $3 \times 10^{24}$ g for a 1 AU disk with $\delta_{\rm 20\, AU} = 8 \times 10^{-3}$. Limits on the CO mass derived from 4.7 $\mu$m CO rovibrational lines (\citealt{Brown12}, Figure 10) are highly dependent on the gas temperature and vary between $10^{30}$ g for 100 K and $10^{19}$ g for 1000 K. For gas temperatures $>$300 K, predicted by our model, better constraints on the CO mass inside the 20 AU gas hole can be obtained from the rovibrational lines ($<10^{21}$ g).

\subsection{Prediction for the CO isotopologue lines} \label{sec:coisotop}
 
\begin{figure*}[htb!]
\center
\includegraphics[width=1.0\hsize]{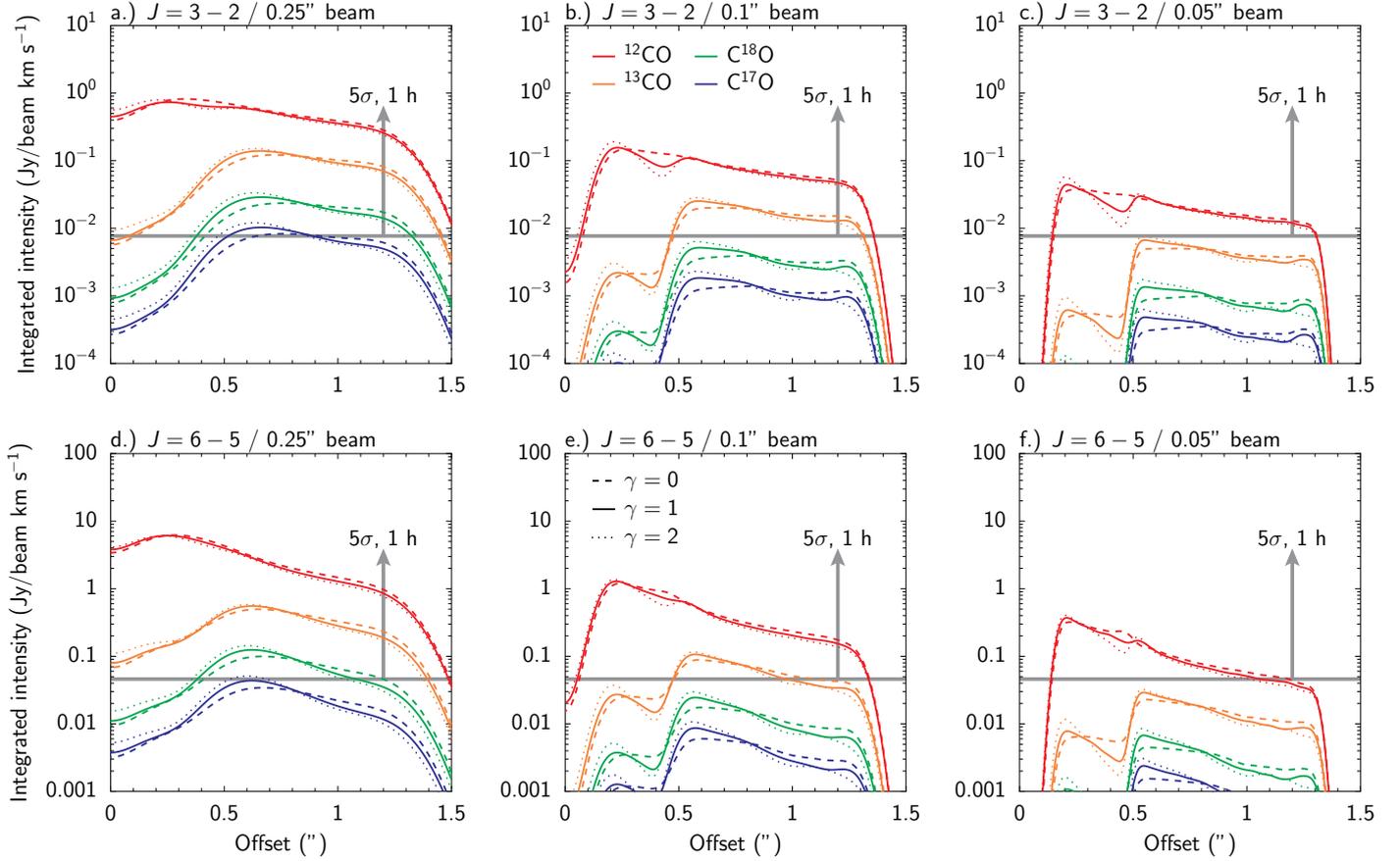}
\caption{Beam convolved integrated emission of the CO isotopologues for the $J=3-2$ and $J=6-5$ transition shown along the cut through the major axis of the disk. The cuts are given for beam sizes of 0.25'', 0.1'' and 0.05''. The $5\sigma$ detection limit within 1 h of the full ALMA using 50 12 m antennas is indicated by the grey horizontal line. Models with a surface density profile slope of $\gamma=0$ (dashed line), $\gamma=1$ (solid line), and $\gamma=2$ (dotted line) are shown.}
\label{fig:iso}
\end{figure*}

\begin{figure*}[htb!]
\center
\includegraphics[width=1.0\hsize]{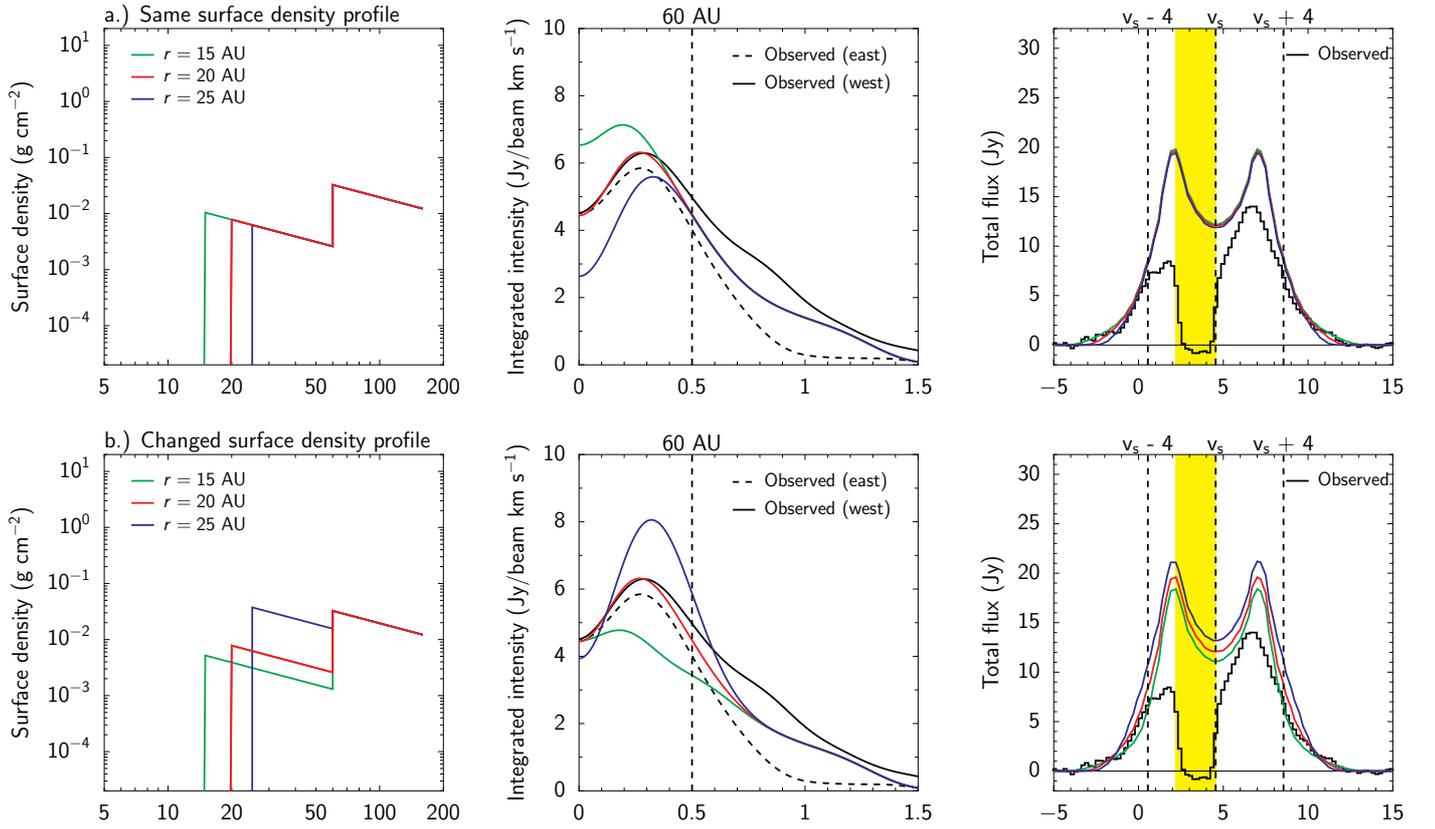}
\caption{Comparison of modeled and observed $^{12}$CO $J=6-5$ intensity profiles and spectra, showing the effect of changing the inner radius of the gas disk at around 20 AU.\textit{Left panels:} Adopted surface density profiles with variations of the representative model. The red line is the representative model. \textit{Center panel:} Intensity profile extracted along the major axis of the disk. The dashed vertical line indicates the radius of the dust free cavity. \textit{Right panel:} Total spectrum. The yellow region indicates regions affected by the foreground. Vertical dashed lines show the source velocity ${\rm v}_{\rm s}=4.55$ km s$^{-1}$ and velocities ${\rm v}_{\rm s} \pm 4$ km s$^{-1}$ representing velocity ranges corresponding to the inner disk (radii $\lesssim 60$ AU). \textit{a)} Using the gas surface density profile of the representative model. \textit{b)} By varying the surface density such that the emission at the stellar position is approximately reproduced.}\label{fig:surfcomp2}
\end{figure*}

\begin{figure*}[htb!]
\center
\includegraphics[width=1.0\hsize]{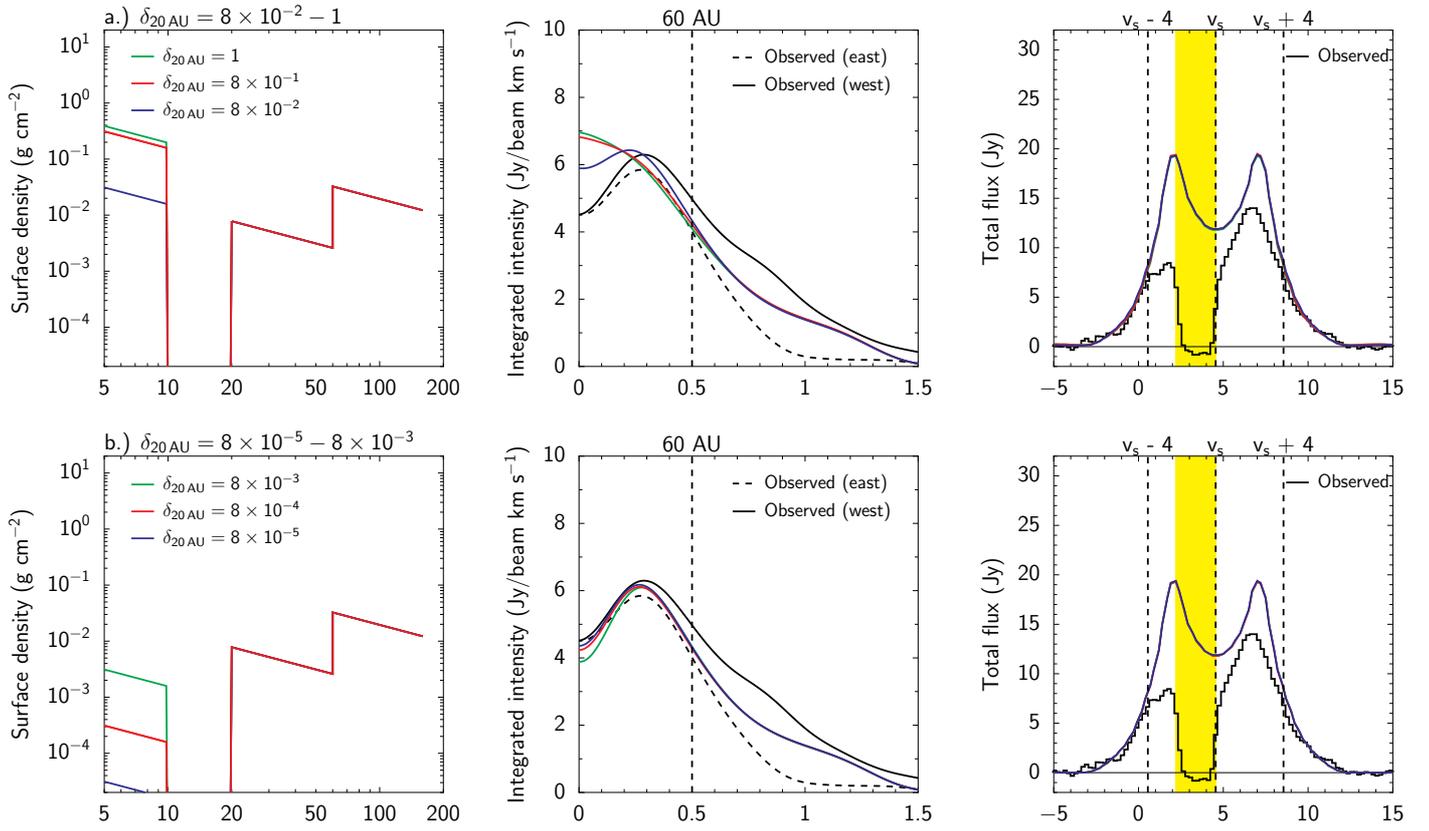}
\caption{Comparison of modeled and observed $^{12}$CO $J=6-5$ intensity profiles and spectra, showing the effect of an inner gas disk extending from 0.4-10 AU. The surface density profile outside 20 AU is set to the representative model.\textit{Left panels:} Adopted surface density profiles with variations of the representative model. The red line is the representative model. \textit{Center panel:} Intensity profile extracted along the major axis of the disk. The dashed vertical line indicates the radius of the dust free cavity. \textit{Right panel:} Total spectrum. The yellow region indicates regions affected by the foreground. Vertical dashed lines show the source velocity ${\rm v}_{\rm s}=4.55$ km s$^{-1}$ and velocities ${\rm v}_{\rm s} \pm 4$ km s$^{-1}$ representing velocity ranges corresponding to the inner disk (radii $\lesssim 60$ AU). \textit{a)} Adopting $\delta_{\rm 20 AU}=8\times 10^{-2} - 1$. \textit{b)} Adopting $\delta_{\rm 20 AU}=8\times 10^{-5} - 8\times 10^{-3}$.}\label{fig:surfcomp3}
\end{figure*}

\begin{figure*}[htb!]
\sidecaption
\includegraphics[width=0.7\hsize]{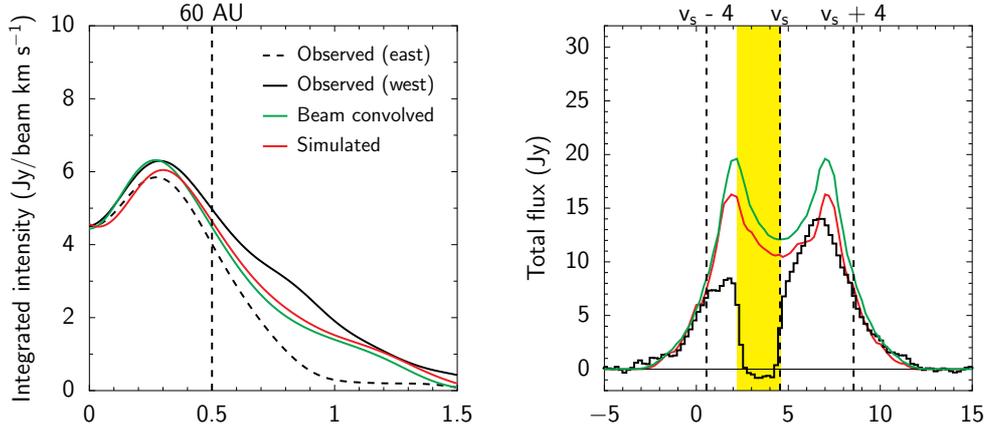}
\caption{Comparison of the intensity profile and total spectrum of the representative model convolved to the ALMA beam or simulated using the observed ${\rm (u,v)}$-coverage. \textit{Left panel:} Intensity profile extracted along the major axis of the disk. The dashed vertical line indicates the radius of the dust free cavity. \textit{Right panel:} Total spectrum. The yellow region indicates regions affected by the foreground. Vertical dashed lines show the source velocity ${\rm v}_{\rm s}=4.55$ km s$^{-1}$ and velocities ${\rm v}_{\rm s} \pm 4$ km s$^{-1}$ representing velocity ranges corresponding to the inner disk (radii $\lesssim 60$ AU).}\label{fig:simul}
\end{figure*}

\begin{figure*}[htb!]
\center
\includegraphics[width=1.0\hsize]{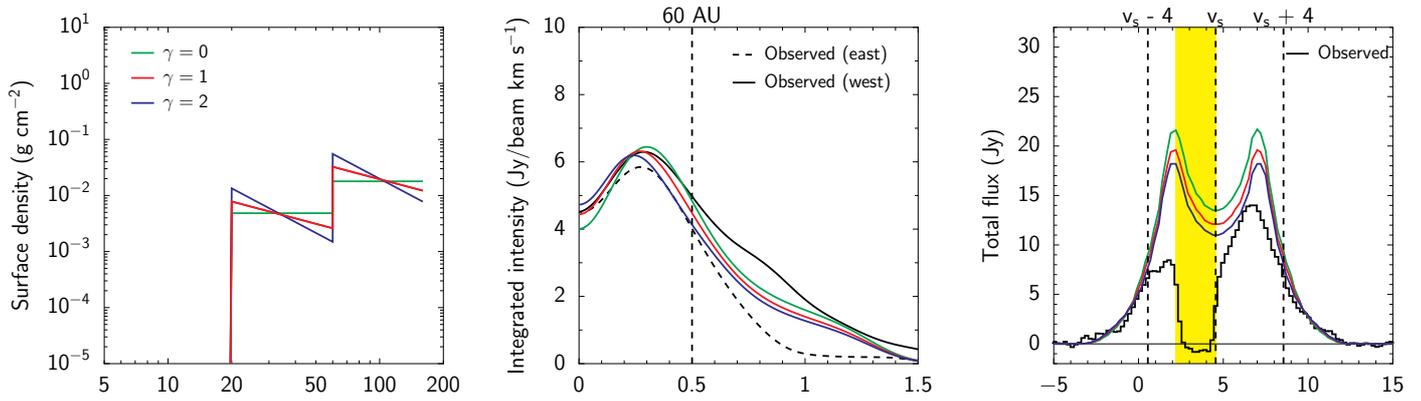}
\caption{Comparison of modeled and observed $^{12}$CO $J=6-5$ intensity profiles and spectra, showing the intensity profile and total spectrum of models with different slope $\gamma$ of the gas and dust surface density profile ($\Sigma \propto r^{-\gamma}$). \textit{Left panels:} Adopted surface density profiles with variations of the representative model. The red line is the representative model. \textit{Center panel:} Intensity profile extracted along the major axis of the disk. The dashed vertical line indicates the radius of the dust free cavity. \textit{Right panel:} Total spectrum. The yellow region indicates regions affected by the foreground. Vertical dashed lines show the source velocity ${\rm v}_{\rm s}=4.55$ km s$^{-1}$ and velocities ${\rm v}_{\rm s} \pm 4$ km s$^{-1}$ representing velocity ranges corresponding to the inner disk (radii $\lesssim 60$ AU).}\label{fig:gamma}
\end{figure*}

Figure \ref{fig:iso} presents the integrated intensity cuts of the CO isotopologues $^{12}$CO, $^{13}$CO, C$^{18}$O and C$^{17}$O $J=3-2$ and $J=6-5$ transitions convolved to a beam size (FWHM) of $0.25'', 0.1''$ and $0.05''$ for the representative model. The beam sizes have been chosen to represent the current observations, and ALMA in future Cycles. The detection limit of the full ALMA combining all 50 12 m antennas for a $5 \sigma$ detection with 1 hour on-source observation is 46 mJy km s$^{-1}$ ($J=6-5$) and 7.7 mJy km s$^{-1}$ ($J=3-2$). For 18-21 antennas (current observations), the detection limit is a factor of 2.5 higher. The limits have been calculated with the ALMA sensitivity calculator\footnote{https://almascience.nrao.edu/proposing/sensitivity-calculator}, using a channel width of 0.2 km s$^{-1}$, and an intrinsic line width of 1 km s$^{-1}$. For $J=6-5$, we have assumed good weather conditions (average precipitable water vapor levels PWV$<$0.47 mm) and for $J=3-2$ average weather conditions (PWV$<$0.91 mm).

The Figure shows that for beam sizes of $0.25''$, $^{13}$CO and C$^{18}$O can be easily detected even in this low-mass disk. At the high frequency of the $J=6-5$ transitions, C$^{17}$O is only strong enough for detection when integrated over a larger region as done in Section \ref{sec:result_spec_c17o}. For lower frequency ($J=3-2$), the C$^{17}$O is predicted to be detectable, although only at a 5 $\sigma$ level. For smaller beam sizes than $0.25''$, the detection of the isotopologues is challenging, both for $J=3-2$ and $J=6-5$. The steep drop in integrated intensity at the stellar position for smaller beams shows that high angular resolution is indeed crucial to provide  better constraints on the amount of gas inside 20 AU (Section \ref{sec:analysis_drop20au}). In more typical disks with a higher gas mass, the rare isotopologues can be readily detected by ALMA (\citealt{Bruderer13}).

%%%%%%%%%%%%%%%%%%%%%%%%%%%%%%%%%%%%%%%%%%%%%%%%%%%%%%%%%%%%%%%%%%%%%%%%%%%%%%%%%%%%%%%%%%%%%%%%%
%
% Discussion
%
%%%%%%%%%%%%%%%%%%%%%%%%%%%%%%%%%%%%%%%%%%%%%%%%%%%%%%%%%%%%%%%%%%%%%%%%%%%%%%%%%%%%%%%%%%%%%%%%%
%%%%%%%%%%%%%%%%%%%%%%%%%%%%%%%%%%%%%%%%%%%%%%%%%%%%%%%%%%%%%%%%%%%%%%%%%%%%%%%%%%%%%%%%%%%%%%%%%

\section{Discussion} \label{sec:discussion}

\subsection{The gas mass and surface density profile} \label{sec:disc_gasstruct}

The low total gas mass of IRS 48 with $\sim$1.4 $\times 10^{-4}$ M$_\odot$ compared to the dust mass of $\sim$1.6 $\times 10^{-5}$ M$_\odot$ yields a gas-to-dust ratio of only $\sim$10. Since the dust temperature in this disk around an A-type star is too warm for CO freeze-out (Figure \ref{fig:model_gasmass}) it is not possible that additional gas is missed due to CO being frozen out. In comparison to HD 142527, another disk around a young Herbig star showing a strongly lopsided continuum emission (\citealt{Casassus13,Fukagawa13}), the dust mass of IRS 48 is a factor of 2.7 lower, but this disk also shows indications of a gas-to-dust ratio lower than the ISM value of 100 (\citealt{Fukagawa13}).

What does the low gas-to-dust ratio\footnote{To calculate the gas-to-dust ratio only the observable dust grains with size below a few millimeters are accounted for.} observed towards IRS 48 imply? Since \citet{vdMarel13} found evidence for grain growth, a gas-to-dust ratio larger than 100 would be expected if some of the dust has grown beyond the size of a few millimeters. Such larger grains are not traced by the dust continuum observations. The low gas mass derived in this work thus suggests that some of the gas is removed from the disk. 

One possible mechanism for gas removal is a wind, as driven by photoevaporation. Photoevaporation has a short processing time-scale and would quickly remove all gas inside the dust cavity (e.g. \citealt{Clarke01}). In IRS 48, however, we find a decrease in the gas surface density at 60 AU of only a factor of $\sim$10, thus making photoevaporation as the only mechanism forming the 60 AU cavity very unlikely, given the short processing time-scale. The drop at 20 AU may not be constrained by our observations and could be deep. However, at least X-ray driven photoevaporation models (\citealt{Owen11,Rosotti13}) fail to explain hole sizes of 20 AU for an accretion rate of $4 \times 10^{-9}$ M$_\odot$ yr$^{-1}$ measured towards IRS 48 (\citealt{Salyk13}). This accretion rate is also higher than typical EUV photoevaporation rates (\citealt{Alexander06a,Alexander06}) ruling out EUV driven evaporation. We thus conclude that in IRS 48 photoevaporation is unlikely the reason for the gap formation at 20 and 60 AU.

Another mechanism leading to a lower gas-to-dust ratio is, that gas is bound into one or several giant planets or companions in the dust depleted region inside 60 AU. The dust trap found towards IRS 48, preventing the dust to drift inwards, could filter the gas from the dust (\citealt{Zhu12}). Assuming an initial gas-to-dust ratio of 100, the amount of gas that has been removed corresponds to $\sim$1.5 M$_{\rm Jupiter}$, which can easily be bound into one companion or giant planet. The companion or giant planet is however not limited to this mass if the initial gas-to-dust mass ratio was 100, because some of the dust could currently be bound in grains too large to be observed by continuum emission. It is interesting to note that our proposed surface density profile combined with the calculated temperature structure leads to a positive pressure gradient ($dp/dr > 0$) at 60 AU as required to trap the dust radially. This positive pressure gradient is found independent of the assumed surface density slope $\gamma$.

\subsection{The gas surface density at the dust trap} \label{sec:disc_surfdusttrap}

The current observational data do not allow us to directly determine the surface density at the position of the dust trap (\citealt{vdMarel13}), since the $^{12}$CO emission is optically thick and the C$^{17}$O emission too weak for mapping.  The difference of the $^{12}$CO integrated intensity at the dust trap compared to the same position in the north by a factor of $\sim$3 is quite large (Section \ref{sec:result_int}). One option to explain this would be a gas temperature difference.

The comparison of a model with gas temperature fixed to the dust temperature to that with calculated gas temperature shows that the $^{12}$CO $J=6-5$ emission in the outer disk ($>$60 AU) emerges from a layer where gas and dust temperature are coupled ($T_{\rm gas}=T_{\rm dust}$). Thus, in order to decrease the $^{12}$CO emission at the dust trap by a factor of 3, the dust temperature needs to be lower by a factor of 3, which in turn implies a factor of about $3^4=81$ less (continuum) radiation should be intercept by the dust, possibly due to shielding by an asymmetric inner disk. The UV pumped CO rovibrational lines observed by \citet{Brown12} show indeed a strong north/south gradient in intensity, but the VLT-VISIR image by \citet{Geers07} (Figure \ref{fig:visir_irs48}) indicates that stellar light is received by the south part of the disk. It is thus unclear how a factor of 3 lower dust temperature in the south can be explained. Since the interpretation of the dust trap requires an overdensity of a factor of $2-3$ at the position of the continuum emission, the temperature decrease should actually be even larger. The time-scale to smooth-out dust asymmetries is of the order of Myrs (\citealt{Birnstiel13}), and it could also be that the gas asymmetry which created the dust trap has already disappeared. We conclude that unless the layer where gas and dust temperatures decouple is substantially larger than determined by our models, a temperature difference of a factor of 3 is difficult to explain. Other scenarios for the observed asymmetry based on future observations of optically thin line emission need to be explored. Multiple transition observations in order to determine the gas temperature directly will facilitate this exploration.

\subsection{Consequences of the gas mass for a companion} \label{sec:disc_masscomp}

Assuming that the 20 AU cavity is carved out by a companion or a giant planet, can we derive constraints on the mass of the companion from the proposed gas surface density profile? For a detailed analysis, hydrodynamical models need to be run in which the depth of the gap in the model is compared to that observed, which is beyond the scope of the study. We only note here that a one Jupiter mass planet is already able to carve out a gap with a depth of a factor of 10 to 1000, depending on the viscosity parameter (\citealt{Pinilla12b}). For higher masses, the gravitational torque is more important than the viscous torque and the depth of the gap carved out by a 9 $M_{\rm Jupiter}$ mass is about a factor of 1000, independent of viscosity. Relating the drop in surface density to the planet mass thus requires additional constraints on the viscosity, which are observationally difficult to obtain. In order not to stir up the gas too much and allow dust to be azimuthally trapped, \citet{vdMarel13} propose a low viscosity parameter $\alpha_t = 10^{-4}$. With such a low $\alpha_t$, already a giant planet with 1 Jupiter mass could carve out the upper limit on the gap depth at 20 AU in IRS 48.

Hydrodynamical models of planets embedded in disks predict a gap opening as far out as $\sim$2 times the orbital radius of the planet (e.g. \citealt{Pinilla12b}). A companion situated at 20 AU is thus unlikely to create the observed drop in surface density of a factor $>$10 at 60 AU. Possibly, IRS 48 harbors a multiple system as proposed for other transitional disks (\citealt{DodsonRobinson11}, \citealt{Zhu11}) with a more massive planet inside 20 AU and one or several less massive planets between 20 and 60 AU. More generally, the drops in the gas density structure such as derived here are potentially the best indicators of just formed planets that are too faint to detect directly.

%%%%%%%%%%%%%%%%%%%%%%%%%%%%%%%%%%%%%%%%%%%%%%%%%%%%%%%%%%%%%%%%%%%%%%%%%%%%%%%%%%%%%%%%%%%%%%%%%
%
% Conclusions
%
%%%%%%%%%%%%%%%%%%%%%%%%%%%%%%%%%%%%%%%%%%%%%%%%%%%%%%%%%%%%%%%%%%%%%%%%%%%%%%%%%%%%%%%%%%%%%%%%%
%%%%%%%%%%%%%%%%%%%%%%%%%%%%%%%%%%%%%%%%%%%%%%%%%%%%%%%%%%%%%%%%%%%%%%%%%%%%%%%%%%%%%%%%%%%%%%%%%

\section{Conclusions}

In this work, we have analyzed high-resolution ALMA submillimeter observations of $^{12}$CO and C$^{17}$O $J=6-5$ rotational emission from a transition disk using a chemical-physical model. With the model we can set constraints on the surface density profile and the total gas mass of the disk. We can furthermore set constraints on the gas inside a 20 AU cavity and in the 20-60 AU gap, which are thought to be carved out by a companion or one or more giant planets. The main conclusions of our work are:
\begin{itemize}
\item The IRS 48 disk has a much lower gas mass than previously derived ($\sim$1.4$\times 10^{-4}$ M$_\odot$). Together with the dust mass of $\sim$1.6$\times 10^{-5}$ M$_\odot$, the gas mass leads to a gas-to-dust ratio of only 10. Given the uncertainties in determining the gas and dust mass of both a factor of a few, this gas-to-dust ratio is however only marginally below the ISM value of 100. The low gas-to-dust ratio is possibly related to gas filtered from dust at the position of the dust trap, combined with gas inside the dust trap bound into a companion or giant planet.
\item Inside a radius of 60 AU, where small dust is thought to be absent, we find that a decrease of the gas surface density profile of a factor of $\sim$12 is required to reproduce observations. Inside a radius of 20 AU, where no CO rotational emission is detected, a drop of at least an  additional factor of $\gtrsim$10 is derived, in total a drop of a factor of $\gtrsim$110. To set more stringent constraints, higher resolution observations are required. Observations with beam size of order $0.1''$ are predicted to be sufficient to constrain the drop considerably better. The drops at 20 AU and 60 AU are derived assuming a slope $\gamma=1$ of the surface density profile ($\Sigma \propto r^{-\gamma}$ ). For $\gamma=0$ and $\gamma=2$, which are also consistent with the data, the drops are a factor of 3 smaller/larger.
\item The inner radius of gas depletion at 20 AU can be constrained to better than $\pm$5 AU.
\item Fitting the SED and the 18.7 $\mu$m dust continuum image taken by VLT-VISIR requires a small ($<$1 AU) dusty disk close to the star and a large scale height of the outer disk (scale-height angle $h_{\rm 60\, AU}=0.14$ radians).
\item Despite of the low gas-mass of IRS 48, the full ALMA combining all 50 antennas is predicted to readily detect $^{13}$CO $J=6-5$ and C$^{18}$O $J=6-5$. These optically thin isotopologue lines will help to better constrain the gas surface density profile of the disk and to study different scenarios of azimuthal asymmetries. 
\item The derived gas surface density profile points to the clearing of the cavity by one or several massive planets or companions rather than just photoevaporation or grain-growth.
\end{itemize}

\begin{acknowledgements}
This paper makes use of the following ALMA data: ADS/JAO.ALMA no. 2011.0.00635.SSB. ALMA is a partnership of the European Southern Observatory (ESO) (representing its member states), NSF (USA), and National Institutes of Natural Sciences (Japan), together with the National Research Council (Canada) and National Science Council and Academia Sinica Institute of Astronomy and Astrophysics (Taiwan), in cooperation with the Republic of Chile. The Joint ALMA Observatory is operated by the ESO, Associated Universities Inc./ National Radio Astronomy Observatory, and National Astronomical Observatory of Japan. We thank an anonymous referee, Til Birnstiel, Joanna Brown, Cornelis Dullemond, Greg Herczeg, Attila Juh\`asz, Paola Pinilla, Markus Schmalzl, and Malcolm Walmsley for useful comments and discussions. S.B. acknowledges a stipend by the Max Planck Society and N.v.d.M. is supported by the Netherlands Research School for Astronomy (NOVA) Band 9 funding. T.v.K thanks Allegro, funded by NWO Physical Sciences ("the Netherlands Organization for Scientific Research (NWO), Physical Sciences"), for financial and technical support. We thank Melissa McClure for providing the Spitzer data in electronic form. Astrochemistry in Leiden is supported by the Netherlands Research School for Astronomy (NOVA), by a Royal Netherlands Academy of Arts and Sciences (KNAW) professor prize, and by the European Union A-ERC grant 291141 CHEMPLAN.
\end{acknowledgements}

%\bibliographystyle{aa}
%\bibliography{mybib}

\listofobjects

\Online 

\end{document}